# Data Mining the SDSS SkyServer Database

Jim Gray, Don Slutz
Microsoft Research

Alex S. Szalay, Ani R. Thakar, Jan vandenBerg
Johns Hopkins University

Peter Z. Kunszt
CERN

Christopher Stoughton
Fermi National Laboratory

# *Data Mining the SDSS SkyServer Database[1]*
## *Jan 2002*

Jim Gray[1], Alex S. Szalay[2], Ani R. Thakar[2], Peter Z. Kunszt[4], Christopher Stoughton[3], Don Slutz[1], Jan vandenBerg[2]
(1) Microsoft, (2) Johns Hopkins, (3) Fermilab, (4) CERN
Gray@Microsoft.com, drslutz@msn.com, {Szalay, Thakar, Vincent}@pha.JHU.edu, Peter.Kunszt@cern.ch, Stoughto@FNAL.gov

**Abstract:** An earlier paper described the Sloan Digital Sky Survey's (SDSS) data management needs [Szalay1] by defining twenty database queries and twelve data visualization tasks that a good data management system should support. We built a database and interfaces to support both the query load and also a website for ad-hoc access. This paper reports on the database design, describes the data loading pipeline, and reports on the query implementation and performance. The queries typically translated to a single SQL statement. Most queries run in less than 20 seconds, allowing scientists to interactively explore the database. This paper is an in-depth tour of those queries. Readers should first have studied the companion overview paper "The SDSS SkyServer – Public Access to the Sloan Digital Sky Server Data" [Szalay2].

## Introduction

The Sloan Digital Sky Survey (SDSS) is doing a 5-year survey of 1/3 of the celestial sphere using a modern ground-based telescope to about ½ arcsecond resolution [SDSS]. This will observe about 200M objects in 5 optical bands, and will measure the spectra of a million objects.

The raw telescope data is fed through a data analysis pipeline at Fermilab. That pipeline analyzes the images and extracts many attributes for each celestial object. The pipeline also processes the spectra extracting the absorption and emission lines, and many other attributes. This pipeline embodies much of mankind's knowledge of astronomy within a million lines of code [SDSS-EDR]. The pipeline software is a major part of the SDSS project: approximately 25% of the project's total cost and effort. The result is a very large and high-quality catalog of the Northern sky, and of a small stripe of the southern sky. Table 1 summarizes the data sizes. SDSS is a 5 year survey starting in 2000. Each year 5TB more raw data is gathered. The survey will be complete by the end of 2006.

**Table 1**: SDSS data sizes (in 2006) in terabytes. About 7 TB online and 10 TB in archive (for reprocessing if needed).

| Product | Raw | Compressed |
|---|---|---|
| Pipeline input | 25 TB | 10 TB |
| Pipeline output (reduced images) | 10 TB | 4 TB |
| Catalogs | 1 TB | 1 TB |
| Binned sky and masks | ½ TB | ½ TB |
| Atlas images | 1TB | 1TB |

Within a week or two of the observation, the reduced data is available to the SDSS astronomers for validation and analysis. They have been building this telescope and the software since 1989, so they want to have "first rights" to the data. They need great tools to analyze the data and maximize the value of their one-year exclusivity on the data. After a year or so, the SDSS publishes the data to the astronomy community and the public – so in 2007 all the SDSS data will be available to everyone everywhere.

The first data from the SDSS, about 5% of the total survey, is now public. The catalog is about 80GB containing about 14 million objects and 50 thousand spectra. People can access it via the SkyServer (http://skyserver.sdss.org/) on the Internet or they may get a private copy of the data. Amendments to this data will be released as the data analysis pipeline improves, and the data will be augmented as more be-comes public. In addition, the SkyServer will get better documentation and tools as we gain more experience with how it is used.

---

[1] The Alfred P. Sloan Foundation, the Participating Institutions, the National Aeronautics and Space Administration, the National Science Foundation, the U.S. Department of Energy, the Japanese Monbukagakusho, and the Max Planck Society have provided funding for the creation and distribution of the SDSS Archive. The SDSS Web site is http://www.sdss.org/. The Participating Institutions are The University of Chicago, Fermilab, the Institute for Advanced Study, the Japan Participation Group, The Johns Hopkins University, the Max-Planck-Institute for Astronomy (MPIA), the Max-Planck-Institute for Astrophysics (MPA), New Mexico State University, Princeton University, the United States Naval Observatory, and the University of Washington. Compaq donated the hardware for the SkyServer and some other SDSS processing. Microsoft donated the basic software for the SkyServer.

## Database Logical Design

The SDSS processing pipeline at Fermi Lab examines the images from the telescope's 5 color bands and identifies objects as a *star*, a *galaxy*, or other *(trail, cosmic ray, satellite, defect)*. The classification is probabilistic—it is sometimes difficult to distinguish a faint star from a faint galaxy. In addition to the basic classification, the pipeline extracts about 400 object attributes, including a 5-color atlas *cutout* image of the object (the raw pixels).

The actual observations are taken in *stripes* that are about 2.5° wide and 130° long. The stripes are processed one *field* at a time (a field has 5 color *frames* as in figure 2.) Each field in turn contains many *objects*. These stripes are in fact the mosaic of two night's observation (two *strips*) with about 10% overlap between the observations. Also, the stripes themselves have some overlaps near the horizon. Consequently, about 10% of the objects appear more than once in the pipeline. The pipeline picks one object instance as *primary* but all instances are recorded in the database. Even more challenging, one star or galaxy often overlaps another, or a star is part of a cluster. In these cases *child* objects are *deblended* from the parent object, and each child also appears in the database (deblended parents are never primary.) In the end about 80% of the objects are primary.

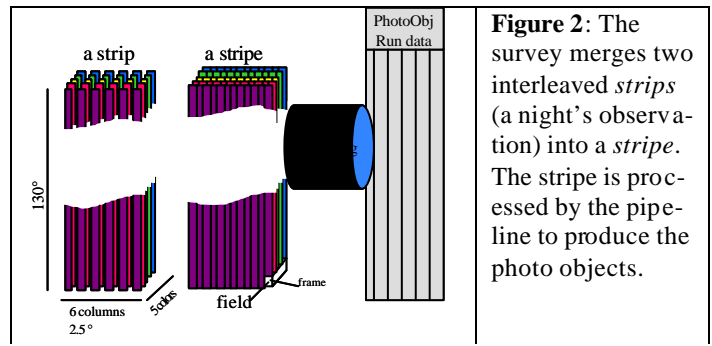

**Figure 2**: The survey merges two interleaved *strips* (a night's observation) into a *stripe*. The stripe is processed by the pipeline to produce the photo objects.

The photo objects have positional attributes (right ascension, declination, (x,y,z) in the J2000 coordinate system, and HTM index). Objects have the five magnitudes and five error bars in five color bands measured in six different ways. Galactic extents are measured in several ways in each of the 5 color bands with error estimates (Petrosian, Stokes, DeVaucouleurs, and ellipticity metrics.) The pipeline assigns a few hundred properties to each object – these attributes are variously called flags, status, and type. In addition to their attributes, objects have a *profile* array, giving the luminance in concentric rings around the object.

The photo object attributes are represented in the SQL database in several ways. SQL lacks arrays or other constructors. So rather than representing the 5 color magnitudes as an array, they are represented as scalars indexed by their names -- ModelMag_r is the name of the "red" magnitude as measured by the best model fit to the data. In other cases, the use of names was less natural (for example in the profile array) and so the data is encapsulated by access functions that extract the array elements from a blob holding the array and its descriptor – for example `array(profile,3,5)` returns `profile[3,5]`. Spectrograms are measured for approximately 1% of the objects. Most objects have estimated (rather than measured) redshifts recorded in the *photoZ* table. To speed spatial queries, a *neighbors* table is computed after the data is loaded. For every object the *neighbors* table contains a list of all other objects within ½ arcminute of the object (typically 10 objects). The pipeline also tries to correlate photo object with objects in other catalogs: United States Naval Observatory [USNO], Röntgen Satellite [ROSAT], Faint Images of the Radio Sky at Twenty-centimeters [FIRST], and others. These correlations are recorded in a set of relationship tables.

The result is a star-schema (see Figure 3) with the *photoObj* table in the center and *fields*, *frames*, *photoZ*, *neighbors*, and connections to other surveys clustered about it. The 14 million *photoObj* records each have about 400 attributes describing the object – about 2KB per record. The *frame* table describes the processing for a particular color band of a field. Not shown in Figure 3 is the metadata *DataConstants* table that holds the names, values, and documentation for all the *photoObj* flags. It allows us to use names rather than binary values (e.g. `flags & fPhotoFlags('primary')`).

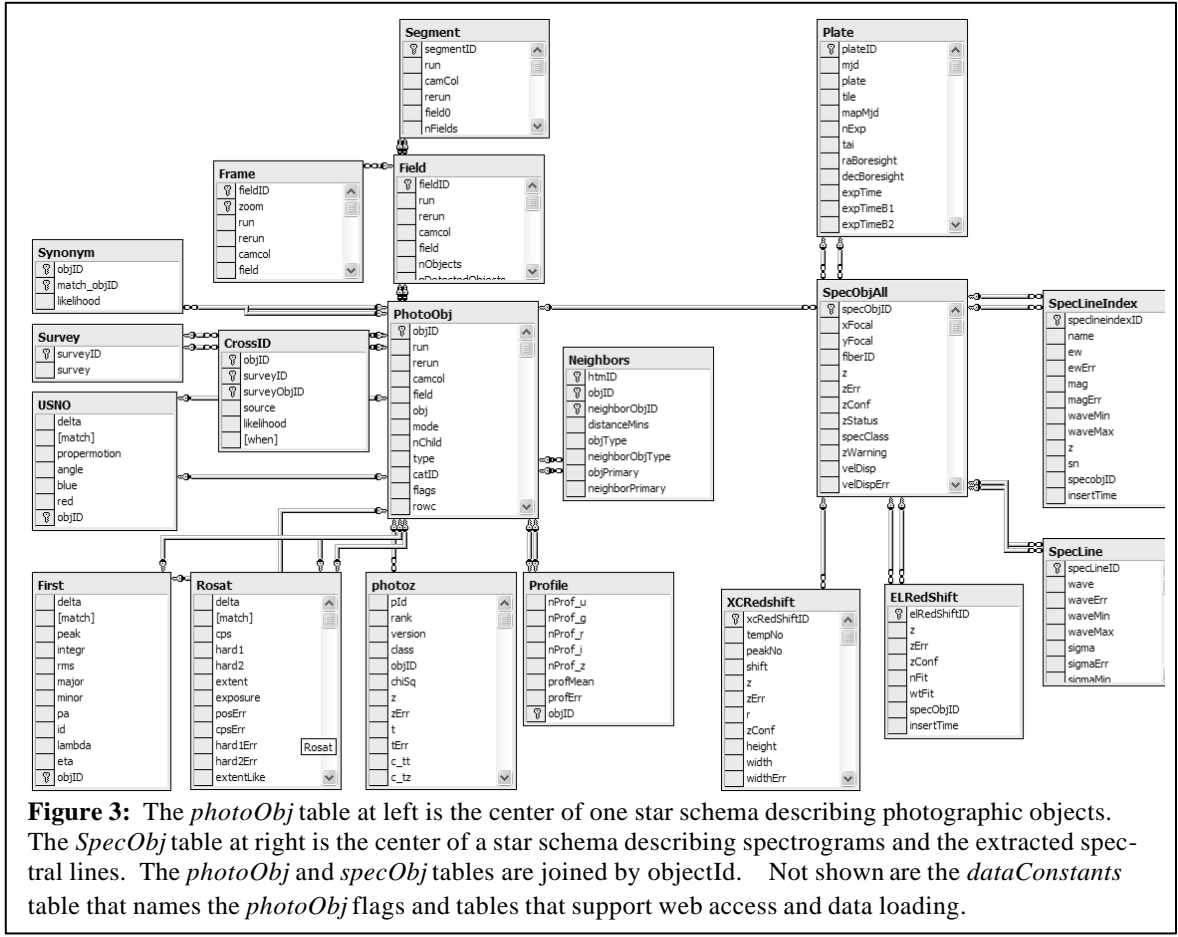

**Figure 3:** The *photoObj* table at left is the center of one star schema describing photographic objects. The *SpecObj* table at right is the center of a star schema describing spectrograms and the extracted spectral lines. The *photoObj* and *specObj* tables are joined by objectId. Not shown are the *dataConstants* table that names the *photoObj* flags and tables that support web access and data loading.

Spectrograms are the second kind of object. About 600 spectra are observed at once using a single *plate* – a metal disk drilled with 600 carefully placed holes, each holding an optical fiber going to a different CCD spectogram. The plate description is stored in the *plate* table, and the description of the spectrogram and its GIF are stored in the *specObj* table. The pipeline processing extracts about 30 spectral lines from each spectrogram. The spectral lines are stored in the *SpecLine* table. The *SpecLineIndex* table has derived line attributes used by astronomers to characterize the types and ages of astronomical objects. Each line is cross-correlated with a model and corrected for redshift. The resulting line attributes are stored in the *xcRedShift* table. Lines characterized as emission lines (about one per spectrogram) are described in the *elRedShift* table.

There is also a set of tables used to monitor the data loading process and to support the web interface. Perhaps the most interesting are the *Tables, Columns, DataConstants,* and *Functions* tables. The SkyServer database schema is documented (in html) as comments in the schema text. We wrote a parser that converts this schema to a collection of tables. Part of the sky server website lets users explore this schema. Having the documentation imbedded in the schema makes maintenance easier and assures that the documentation is consistent with reality (http://skyserver.sdss.org/en/help/docs/browser.asp.) The comments are also presented in *tool tips* by the Query Tool we built

## Database Access Design – Views, Indices, and Access Functions

The *photoObj* table contains many types of objects (primaries, secondaries, stars, galaxies,…). In some cases, users want to see all the objects, but typically, users are just interested in primary objects (best instance of a deblended child), or they want to focus on just Stars, or just Galaxies. Several views are defined on the *PhotoObj* table to facilitate this subset access:

*PhotoPrimary*: *photoObj* records with flags('primary')=true
*PhotoSecondary*: *photoObj* records with flags('secondary')=true
*PhotoFamily*: *photoObj* that is not primary or secondary.
*Sky*: blank sky *photoObj* recods (for calibration).
*Unknown*: *photoObj* records of type "unknown"
*Star*: *PrimaryObjects* subsetted with type='star'
*Galaxy*: *PrimaryObjects* subsetted with type='galaxy'
*SpecObj*: Primary *SpecObjAll* (dups and errors removed)

**Figure 4.** Count of records and bytes in major tables. Indices add 50% more space.

| Table | Records | Bytes |
|---|---|---|
| Field | 14k | 60MB |
| Frame | 73k | 6GB |
| PhotoObj | 14m | 31GB |
| Profile | 14m | 9GB |
| Neighbors | 111m | 5GB |
| Plate | 98 | 80KB |
| SpecObj | 63k | 1GB |
| SpecLine | 1.7m | 225MB |
| SpecLineIndex | 1.8m | 142MB |
| xcRedShift | 1.9m | 157MB |
| elRedShift | 51k | 3MB |

Most users will work in terms of these views rather than the base table. In fact, most of the queries are cast in terms of these views. The SQL query optimizer rewrites such queries so that they map down to the base *photoObj* table with the additional qualifiers.

To speed access, the base tables are heavily indexed (these indices also benefit view access). In a previous design based on an object-oriented database ObjectivityDB™ [Thakar], the architects replicated vertical data slices in *tag* tables that contain the most frequently accessed object attributes. These tag tables are about ten times smaller than the base tables (100 bytes rather than 1,000 bytes) – so a disk-oriented query runs 10x faster if the query can be answered by data in the tag table.

Our concern with the tag table design is that users must know which attributes are in a tag table and must know if their query is "covered" by the fields in the tag table. Indices are an attractive alternative to tag tables. An index on fields A, B, and C gives an automatically managed tag table on those 3 attributes plus the primary key – and the SQL query optimizer automatically uses that index if the query is *covered by* (contains) only those 3 fields. So, indices perform the role of tag tables and lower the intellectual load on the user. In addition to giving a column subset, thereby speeding access by 10x to 100x. Indices can also cluster data so that searches are limited to just one part of the object space. The clustering can be by type (star, galaxy), or space, or magnitude, or any other attribute. Microsoft's SQL Server limits indices to 16 columns – that constrained our design choices.

Today, the SkyServer database has tens of indices, and more will be added as needed. The nice thing about indices is that when they are added, they speed up any queries that can use them. The downside is that they slow down the data insert process – but so far that has not been a problem. About 30% of the SkyServer storage space is devoted to indices.

In addition to the indices, the database design includes a fairly complete set of foreign key declarations to insure that every profile has an object; every object is within a valid field, and so on. We also insist that all fields are non-null. These integrity constraints are invaluable tools in detecting errors during loading and they aid tools that automatically navigate the database. You can explore the database design using web interface at http://skyserver.sdss.org/en/help/docs/browser.asp.

## Spatial Data Access

The SDSS scientists are especially interested in the galactic clustering and large-scale structure of the universe. In addition, the http://skyserver.sdss.org visual interface routinely asks for all objects in a certain rectangular or circular area of the celestial sphere. The SkyServer uses three different coordinate systems. First right-ascension and declination (comparable to latitude-longitude in celestial coordinates) are ubiquitous in astronomy. To make arc-angle computations fast, the (x,y,z) unit vector in J2000 coordinates is stored. The dot product or the Cartesian difference of two vectors are quick ways to determine the arc-angle or distance between them.

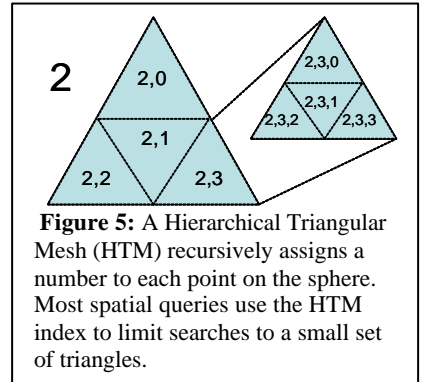

**Figure 5:** A Hierarchical Triangular Mesh (HTM) recursively assigns a number to each point on the sphere. Most spatial queries use the HTM index to limit searches to a small set of triangles.

To make spatial area queries run quickly, we integrated the Johns Hopkins *hierarchical triangular mesh* (HTM) code [HTM, Kunszt] with SQL Server. Briefly, HTM inscribes the celestial sphere within an octahedron and projects each celestial point onto the surface of the octahedron. This projection is approximately iso-area. The 8 octahedron triangular faces are each recursively decomposed into 4 sub-triangles. SDSS uses a 20-deep HTM so that the individual triangles are less than .1 square arcsecond.

The HTM ID for a point very near the north pole (in galactic coordinates) would be something like 2,3,,3 (see Figure 5). These HTM IDs are encoded as 64-bit strings (bigints). Importantly, all the HTM IDs within the triangle 6,1,2,2 have HTM IDs that are between 6,1,2,2 and 6,1,2,3. When the HTM IDs are stored in a B-tree index, simple range queries provide quick index for all the objects within a given triangle.

The HTM library is an external stored procedure wrapped in a table-valued stored procedure spHTM_Cover(<area>). The <area> can be either a circle (ra, dec, radius), a half-space (the intersection of planes), or a polygon defined by a sequence of points. A typical area might be 'CIRCLE J2000, 30.1, -10.2 .8' which defines an 0.8 arc minute circle around the (ra,dec) = (30.1, -10.2)[2]. The spHTM_Cover table valued function has the following template:

```
CREATE FUNCTION spHTM_Cover (@Area VARCHAR(8000))    -- the area to cover
    RETURNS @Triangles TABLE (                        -- returns table
        HTMIDstart BIGINT NOT NULL PRIMARY KEY,       -- start of triangle
        HTMIDend   BIGINT NOT NULL)                   -- end of triangle
```

The procedure call: `select * from spHTM_Cover('Circle J2000 12 5.5 60.2 1')` returns the following table with four rows, each row defining the start and end of a 12-deep HTM triangle.

| HTMIDstart | HTMIDend |
|---|---|
| 3,3,2,0,0,1,0,0,1,3,2,2,2,0 | 3,3,2,0,0,1,0,0,1,3,2,2,2,1 |
| 3,3,2,0,0,1,0,0,1,3,2,2,2,2 | 3,3,2,0,0,1,0,0,1,3,2,2,3,0 |
| 3,3,2,0,0,1,0,0,1,3,2,3,0,0 | 3,3,2,0,0,1,0,0,1,3,2,3,1,0 |
| 3,3,2,0,0,1,0,0,1,3,2,3,3,1 | 3,3,2,0,0,1,0,0,1,3,3,0,0,0 |

One can join this table with the *photoObj* or *specObj* tables to get spatial subsets. There are many examples of this in the sample queries below (see Q1 for example).

The `spHTM_Cover()` function is a little too primitive for most users, they actually want the objects nearby a certain object, or they want all the objects in a certain area – and they do not want to have to pick the HTM depth. So, the following family of functions is supported:

```
fGet{Nearest | Nearby} {Obj | Frame | Mosaic} Eq  (ra, dec, radius_arc_minutes)
fGet{Nearest | Nearby} {Obj | Frame | Mosaic} XYZ (x, y, z, radius_arc_minutes)
```

---

[2] The full syntax for areas is:

```
CIRCLE     J2000       depth ra dec radius_arc_minutes
CIRCLE     CARTESIAN   depth x y z radius_arc_minutes
CONVEX     J2000       depth n ra1 dec1 ra2 dec2 ..... ran decn              // a polygon
CONVEX     CARTESIAN   x1 y1 z1 x2 y2 z2..... xn yn zn                        // a polygon
DOMAIN     depth  k    n1 x1 y1 z1 d1 x2 y2 z2 d2... xn1 yn1 zn1 dn1
                       n2 x1 y1 z1 d1 x2 y2 z2 d2... xn2 yn2 zn2 dn2
                       ..
                       nk  x1 y1 z1 d1 x2 y2 z2 d2... xnk ynk znk dnk
```

For example: `fGetNeaestObjEq(1,1,1)` returns the nearest object coordinates within one arcminute of equatorial coordinate (1º, 1º). These procedures are frequently used in the 20 queries and in the website access pages.

In summary, the logical database design consists of photographic and spectrographic objects. They are organized into a pair of snowflake schema. Subsetting views and many indices give convenient access to the conventional subsets (stars, galaxies, ...). Several procedures are defined to make spatial lookups convenient. http://skyserver.sdss.org/en/help/docs/browser.asp documents these functions in more detail.

## Database Physical Design and Performance

The SkyServer initially took a simple approach to database design – and since that worked, we stopped there. The design counts on the SQL Server data storage engine and query optimizer to make all the intelligent decisions about data layout and data access.

The data tables are all created in one file group. The file group consists of files spread across all the disks. If there is only one disk, this means that all the data (about 80 GB) is on one disk, but more typically there are 4 or 8 disks. Each of the *N* disks holds a file that starts out as size 80 GB*/N* and automatically grows as needed. SQL Server stripes all the tables across all these files and hence across all these disks. When reading or writing, this automatically gives the sum of the disk bandwidths without any special user programming. SQL Server detects the sequential access, creates the parallel prefetch threads, and uses multiple processors to analyze the data as quickly as the disks can produce it. Using commodity low-end servers we measure read rates of 150 MBps to 450 MBps depending on how the disks are configured.

Beyond this file group striping; SkyServer uses all the SQL Server default values. There is no special tuning. This is the hallmark of SQL Server – the system aims to have "no knobs" so that the out-of-the box performance is quite good. The SkyServer is a testimonial to that goal.

So, how well does this work? The appendix gives detailed timings on the twenty queries; but, to summarize, a typical index lookup runs primarily in memory and completes within a second or two. SQL Server expands the database buffer pool to cache frequently used data in the available memory. Index scans of the 14M row photo table run in 7 seconds "warm" (2 m records per second when CPU-bound), and 18 seconds cold (100 MBps when disk bound), on a 4-disk 2-CPU Server. Queries that scan the entire 30 GB *photoObj* table run at about 150MBps and so take about 3 minutes. These scans use the available CPUs and disks to run in parallel. In general we see 4-disk workstation-class machines running at the 150 MBps, while 8-disk server-class machines can run at 300 MBps.

When the SkyServer project began, the existing software (ObjectivityDB™ on Linux or Windows) was delivering 0.5 MBps and heavy CPU consumption. That performance has now improved to 300 MBps and about 20 instructions per byte (measured at the SQL level). This gives 5-second response to simple queries, and 5-minute response to full database scans. The SkyServer goal was 50MBps at the user level on a single machine. As it stands SQL Server and the Compaq hardware exceeded these performance goals by 500% -- so we are *very* pleased with the design. As the SDSS data grows, arrays of more powerful machines should allow the SkyServer to return most answers within seconds or minutes depending on whether it is an index search, or a full-database scan.

## Database Load Process

The SkyServer is a data warehouse: new data is added in batches, but mostly the data is queried. Of course these queries create intermediate results and may deposit their answers in temporary tables, but the vast bulk of the data is read-only.

Occasionally, a brand new schema must be loaded, so the disks were chosen to be large enough to hold three complete copies of the database (70GB disks).

From the SkyServer administrator's perspective, the main task is data loading -- which includes data validation. When new photo objects or spectrograms come out of the pipeline, they must be added to the da-tabase quickly. We are the system administrators – so we wanted this loading process to be as automatic as possible.

The Beowulf data pipeline produces FITS files [FITS]. A filter program converts this output to produce column-separated values (CSV) files, and PNG files [SDSS-EDR]. These files are then copied to the SkyServer. From there, a script-level utility we wrote loads the data using the SQL Server's Data Transformation Service (DTS). DTS does both data conversion and the integrity checks. It also recognizes file names in some fields, and uses the name to insert the image file (PNG or JPEG) as a blob field of the record. There is a DTS script for each table load step. In addition to loading the data, these DTS scripts write records in a *loadEvents* table recording the time of the load, the number of records in the source file and the number of inserted records. The DTS steps also write trace files indicating the success or errors in the load step. A particular load step may fail because the data violates foreign key constraints, or because the data is invalid (violates integrity constraints.) A web user interface displays the load-events table and makes it easy to examine the CSV file and the load trace file. The operator can (1) undo the load step, (2) diagnose and fix the data problem, and (3) re-execute the load on the corrected data. If the input file is easily repaired, that is done by the administrator, but often the data needs to be regenerated. In either case the first step is to UNDO the failed load step. Hence, the web interface has an UNDO button for each step.

The UNDO function works as follows. Each table in the database has an additional timestamp field that records when the record was inserted (the field has `Current_Timestamp` as its default value.) The load event record records the table name and the start and stop time of the load step. Undo consists of deleting all records from the target table with an insert time between that start and stop time.

Loading runs at about 5 GB per hour (data conversion is very CPU intensive), so the current SkyServer loads in about 12 hours. More than ½ this time goes into building or maintaining the indices.

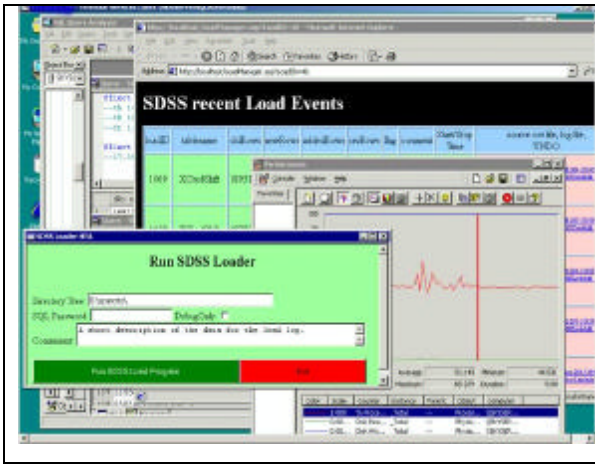

**Figure 6:** A screen shot of the SkyServer Database operations interface. The SkyServer is operated via the Internet using Windows™ Terminal Server, a remote desktop facility built into the operating system. Both loading and software maintenance are done in this way. This screen shot shows a window into the backend system after a load step has completed. It shows the loader utility, the load monitor, a performance monitor window and a database query window. This remote operation has proved a godsend, allowing the Johns Hopkins, Microsoft, and Fermi Lab participants to perform operations tasks from their offices, homes, or hotel rooms.

## Personal SkyServer

A 1% subset of the SkyServer database (about 1/2 GB) that can fit on a CD or downloaded over the web (http://research.microsoft.com/~Gray/PersonalSkyServerV3.zip.) This includes the web site and all the photo and spectrographic objects in a 6º square of the sky. This personal SkyServer fits on laptops and desktops. It is useful for experimenting with queries, for developing the web site, and for giving demos. We also believe SkyServer will be great for education --teaching both how to build a web site and how to do computational science. Essentially, any classroom can have a mini-SkyServer per student. With disk technology improvements, a large slice of the public data will fit on a single disk by 2003.

## Hardware Design and Raw Performance

The SkyServer database is about 80 GB. It can run on a single processor system with just one disk, but the production SkyServer runs on more capable hardware generously donated by Compaq Computer Corporation. Figure 7 shows the hardware configuration.

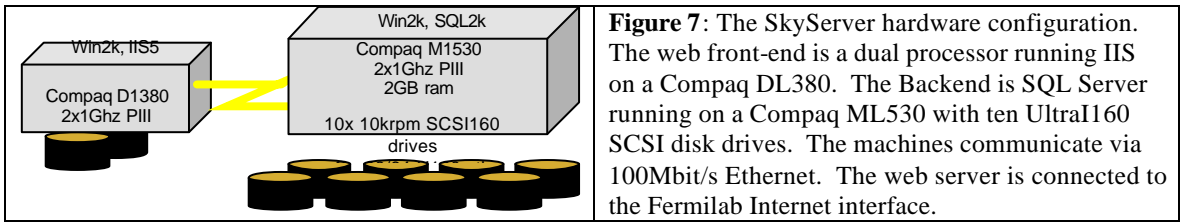

**Figure 7**: The SkyServer hardware configuration. The web front-end is a dual processor running IIS on a Compaq DL380. The Backend is SQL Server running on a Compaq ML530 with ten UltraI160 SCSI disk drives. The machines communicate via 100Mbit/s Ethernet. The web server is connected to the Fermilab Internet interface.

The web server runs Windows2000 on a Compaq ProLiant DL380 with dual 1GHz Pentium III processors. It has 1GB of 133MHz SDRAM, and two mirrored Compaq 37GB 10K rpm Ultra160 SCSI disks attached to a Compaq 64-Bit/66MHz Single Channel Ultra3 SCSI Adapter. This web server does almost no disk IO during normal operation, but we clocked the disk subsystem at over 30MB/s. The web server is also a firewall, it does not do routing and so acts as a firewall. It has a separate "private" 100Mbit/s Ethernet link to the backend database server.

Most data mining queries are IO-bound, so the database server is configured to give fast sequential disk bandwidth. It also helps to have healthy CPU power and high availability. The database server is a Compaq ProLiant ML530 running SQL Server 2000 and Windows2000. It has two 1GHz Pentium III Xeon processors, 2GB of 133MHz SDRAM, a 2-slot 64bit/66MHz PCI bus, a 5-slot 64bit/33MHz PCI bus, and a 32bit PCI bus with a single expansion slot. It has 12 drive bays for low-profile (1 inch) hot-pluggable SCA-2 SCSI drives, split into two SCSI channels of six disks each. It has an onboard dual-channel ultra2 LVD SCSI controller, but we wanted greater disk bandwidth, so we added two Compaq 64-Bit/66MHz Single Channel Ultra3 SCSI Adapters to the 64bit/66MHz PCI bus, and left the onboard ultra2 SCSI controller disconnected. These Compaq ultra160 SCSI adapters are Adaptec 29160 cards with a Compaq BIOS.

The DL380 and the ML530 also have a complement of high-availability hardware components: redundant hot-swappable power supplies, redundant hot-swappable fans, and hot-swappable SCA-2 SCSI disks.

The production database server is configured with 10 Compaq 37GB 10K rpm Ultra160 SCSI disks, five on each SCSI channel. We use Windows 2000's native software RAID to manage the disks as five mirrors (RAID1's), with each mirror split across the two SCSI channels. One mirrored volume is for the operating system and software, and the remaining four volumes are for database files. The database file groups (data, temp, and log) are spread across these four mirrors. SQL Server stripes the data across the four volumes, effectively managing the data disks as a RAID10 (striping plus mirroring). This configuration can scan data at 140 MB/s for a simple query like:

```
select count(*)
from photoObj
where (r-g)>1.
```

Before the production database server was deployed, we ran some tests to find the maximum IO speed for database queries on our ML530 system. We're quite happy with the 140 MB/s performance of the conservative, reliable production server configuration on the 60GB public EDR (Early Data Release) data. However, we're about to implement an internal SkyServer which will contain about 10 times more data than the public SkyServer: about 500-600GB. For this server, we'll probably need more raw speed.

For the max-speed tests, we used our ML530 system, plus some extra devices that we had on-hand: an assortment of additional 10K rpm ultra160 SCSI disks, a few extra Adaptec 29160 ultra160 SCSI controllers, and an external eight-bay two-channel ultra160 SCSI disk enclosure. We started by trying to find the performance limits of each IO component: the disks, the ultra160 SCSI controllers, the PCI busses, and the memory bus. Once we had a good feel for the IO bottlenecks, we added disks and controllers to test the system's peak performance.

For each test setup, we created a stripe set (RAID0) using Windows 2000's built-in software RAID, and ran two simple tests. First, we used the MemSpeed utility (v2.0 [MemSpeed]) to test raw sequential IO speed using 16-deep unbuffered IOs. MemSpeed issues the IO calls and does no processing on the results, so it gives an idealized, best-case metric. In addition to the unbuffered IO speed, MemSpeed also does several tests on the system's memory and memory bus. It tests memory read, write, and memcpy rates - both single-threaded, and multi-threaded with a thread per system CPU. These memory bandwidth measures suggest the system's maximum IO speed. After running MemSpeed tests, we copied a sample 4GB un-indexed SQL Server database onto the test stripe set and ran a very simple `select count(*)` query to see how SQL Server's performance differed from MemSpeed's idealized results.

Figure 8 shows our performance results.

- **Individual disks:** The tests used three different disk models: the Compaq 10K rpm 37GB disks in the ML530, some Quantum 10K rpm 18GB disks, and a 37GB 10K rpm Seagate disk. The Compaq disks could perform sequential reads at 39.8 MB/s, the old Quantums were the slowest at 37.7 MB/s, and the new Seagate churned out 51.7 MB/s! The "linear quantum" plot on Figure 8 shows the best-case RAID0 performance based on a linear scaleup of our slowest disks.
- **Ultra160 SCSI:** A single ultra160 SCSI channel saturates at about 123 MB/s. It makes no sense to add more than three of disks to a single channel. Ultra160 delivers 77% of its peak advertised 160 MB/s.
- **64bit/33MHz PCI:** With three ultra160 controllers attached to the 64bit/33MHz PCI bus, the bus saturates at about 213 MB/s (80% of its max. burst speed of 267 MB/s). This is not quite enough bandwidth to handle the traffic from six disks.
- **64bit/66MHz PCI:** We didn't have enough disks, controllers, or 64bit/66MHz expansion slots to test the bus's 533 MB/s peak advertised performance.
- **Memory bus:** MemSpeed reported single-threaded read, write, and copy speeds of 590 MB/s, 274 MB/s, and 232 MB/s respectively, and multithreaded read, write, and copy speeds of 849 MB/s, 374 MB/s, and 300 MB/s respectively.

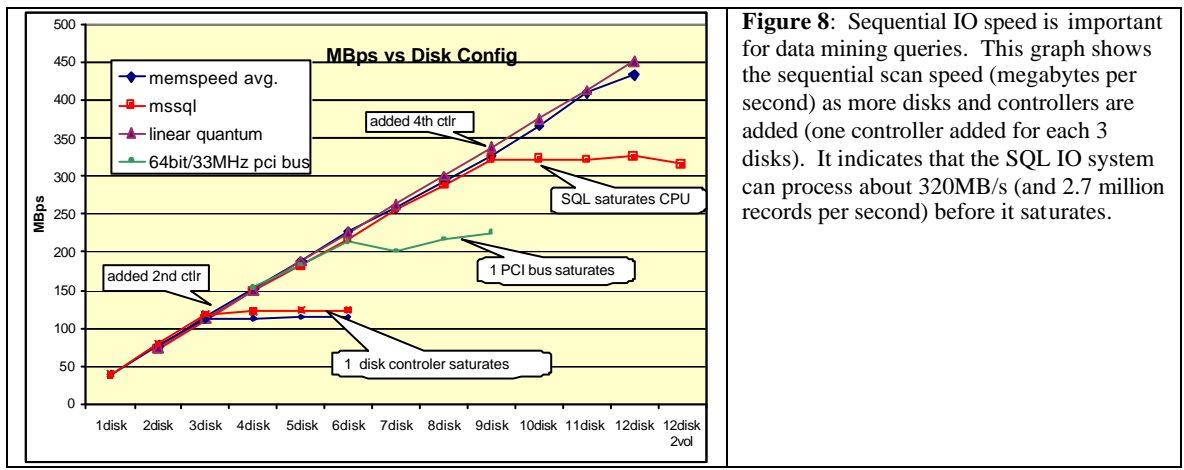

**Figure 8**: Sequential IO speed is important for data mining queries. This graph shows the sequential scan speed (megabytes per second) as more disks and controllers are added (one controller added for each 3 disks). It indicates that the SQL IO system can process about 320MB/s (and 2.7 million records per second) before it saturates.

After the basic component tests, the system was configured to avoid SCSI and PCI bottlenecks. Initially three ultra160 channels were configured: two controllers connected to the 64bit/66MHz PCI bus, and one connected to the 64bit/33MHz bus. Disks were added to the controllers one-by-one, never using more than three disks on a single ultra160 controller. Surprisingly, both the simple MemSpeed tests and the SQL Server tests scaled up linearly almost perfectly to nine disks. The ideal disk speed at nine disks would be 339 MB/s, and we observed 326.7 MB/s from MemSpeed, and 322.4 MB/s from SQL Server. To reach the performance ceiling yet, a fourth ultra160 controller (to the 64bit/33MHz PCI bus) was added along with more disks. The MemSpeed results continued to scale linearly through 11 disks. The 12-disk MemSpeed result fell a bit short of linear at 433.8 MB/s (linear would have been 452 MB/s), but this is probably because we were slightly overloading our 64bit/33MHz PCI bus on the 12-disk test. SQL Server read speed leveled off at 10 disks, remaining in the 322 MB/s ballpark. Interestingly, SQL Server never fully saturated the CPU's for our simple tests. Even at 322 MB/s, CPU utilization was about 85%. Perhaps the memory was saturated at this point. 322 MB/s is in the same neighborhood as the memory write and copy speed limits that we measured with MemSpeed.

Figure 9 shows the relative IO density of the queries. It shows that the queries issue about a thousand IOs per CPU second. Most of these IOs are 64KB sequential reads of the indices or the base data. So, each CPU generates about 64MB of IO per second. Since these CPUs each execute about a billion instructions per second, that translates to an IO density of a million instructions per IO and about 16 instructions per byte of IO – both these numbers are an order of magnitude higher than Amdahl's rules of thumb. Using SQLserver a CPU can consume about five million records per second if the data is in main memory.

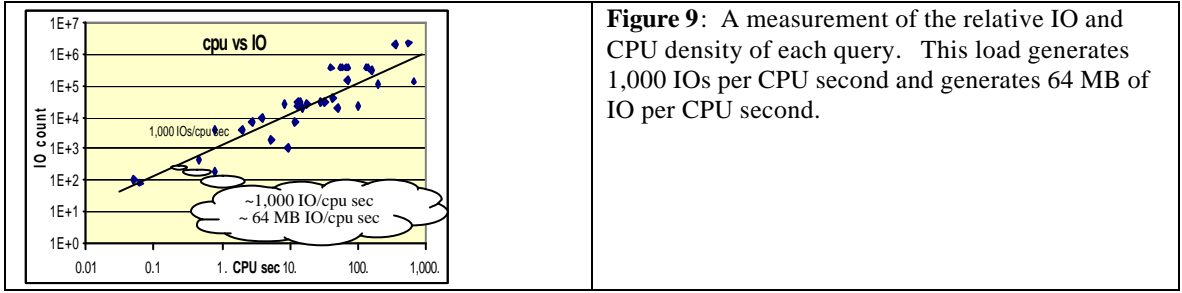

**Figure 9**: A measurement of the relative IO and CPU density of each query. This load generates 1,000 IOs per CPU second and generates 64 MB of IO per CPU second.

## A Summary of the Experience Implementing the Twenty Queries

The Appendix has each of the 20 queries along with a description of the query plans and measurements of the CPU time, elapsed time, and IO demand. This section just summarizes the appendix with general comments.

First, all the 20 queries have fairly simple SQL equivalents. This was not obvious when we started -- and we were very pleased to find it was true. Often the query can be expressed as a single SQL statement. In some cases, the query is iterative, the results of one query feeds into the next. These queries correspond to typical tasks astronomers would do with a TCL script driving a C++ program, extracting data from the archive, and then analyzing it. Traditionally most of these queries would have taken a few days to write in C++ and then a few hours or days to run against the binary files. So, being able to do the query simply and quickly is a real productivity gain for the Astronomy community.

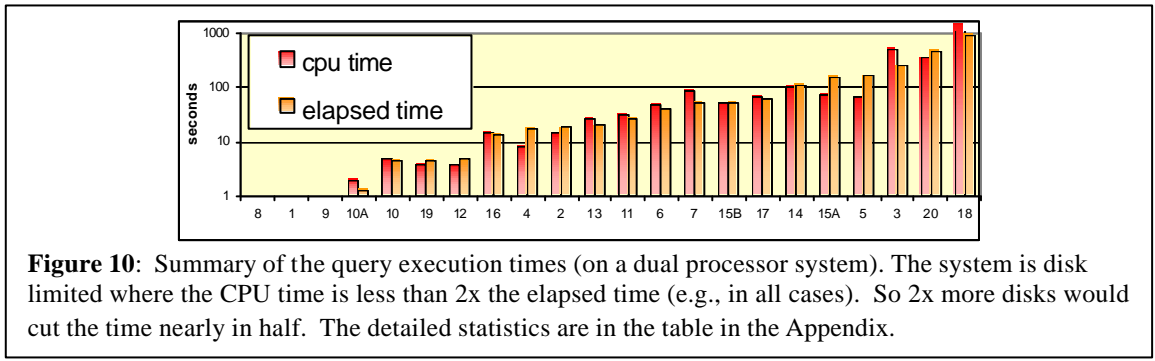

**Figure 10**: Summary of the query execution times (on a dual processor system). The system is disk limited where the CPU time is less than 2x the elapsed time (e.g., in all cases). So 2x more disks would cut the time nearly in half. The detailed statistics are in the table in the Appendix.

Many of the queries run in a few seconds. Some that involve a sequential scan of the database take about 3 minutes. One involves a spatial join and takes ten minutes. As the data grows from 60GB to 1TB, the queries will slow down by a factor of 20. Moore's law will probably give 3x in that time, but still, things will be 7x slower. So, future SkySevers will need more than 2 processors and more than 4 disks. By using CPU and disk parallelism, it should be possible to keep response times in the "few minutes" range.

The spatial data queries are both simple to state and quick to execute using the HTM index. We circumvented a limitation in SQL Server by pre-computing the neighbors of each object. Even without being forced to do it, we would have created this *materialized view* to speed queries. In general, the queries benefited from indices on the popular fields.

In looking at the queries in the Appendix, it is not obvious how they were constructed – they are the finished product. In fact, they were constructed incrementally. First we explored the data a bit to see the rough statistics – either counting (select count(*) from…) or selecting the first 10 answers (select top 10 a,b,c from...). These component queries were then composed to form the final query shown in the Appendix.

It takes both a good understanding of astronomy, a good understanding of SQL, and a good understanding of the database to translate the queries into SQL. In watching how "normal" astronomers access the SX web site, it is clear that they use very simple SQL queries. It appears that they use SQL to extract a subset of the data and then analyze that data on their own system using their own tools. SQL, especially complex SQL involving joins and spatial queries, is just not part of the current astronomy toolkit.

Indeed, our actual query set includes 15 additional queries posed by astronomers using the Objectivity™ archive at (http://archive.stsci.edu/sdss/software/). Those 15 queries are much simpler and run more quickly than most of the original 20 queries.

A good visual query tool that makes it easier to compose SQL would ameliorate part of this problem, but this stands as a barrier to wider use of the SkyServer by the astronomy community. Once the data is pro-duced, there is still a need to understand it. We have not made any progress on the problem of data visualization.

It is interesting to close with two anecdotes about the use of the SkyServer for data mining. First, when it was realized that query 15 (find asteroids) had a trivial solution, one colleague challenged us to find the "fast moving" asteroids (the pipeline detects slow-moving asteroids). These were objects moving so fast, that their detections in the different colors were registered as entirely separate objects (the 5 colors are observed at 5 different 1-minute intervals as the telescope image drifts across the sky – this time-lapse causes slow-moving images to appear as 5 dots of different colors while fast moving images appear as 5 streaks.) This was an excellent test case – our colleague had written a 12 page tcl script that had run for 3 days on the dataset consisting of binary FITS tables. So we had a benchmark to work against. It took a long day to debug our understanding of the data and to develop a query (see query 15A). The resulting query runs in about 10 minutes and finds 3 objects. If we create a supporting index (takes about 10 minutes) then the query runs in less than a minute. Indeed, we have found other fast-moving objects by experimenting with the query parameters. Being able to pose questions in a few hours and get answers in a few minutes changes the way one views the data: you can experiment with it almost interactively. When queries take 3 days and hundreds of lines of code, one asks questions cautiously.

A second story relates to the fact that 99% of the object's spectra will not be measured and so their redshifts will not be measured. As it turns out, the objects' redshifts can be estimated by their 5-color optical measurements. These estimates are surprisingly good [Budavari1, Budavari2]. However, the estimator requires a training set. There was a part of parameter space – where only 3 galaxies were in the training data and so the estimator did a poor job. To improve the estimator, we wanted to measure the spectra of 1,000 such galaxies. Doing that required designing some plates that measure the spectrograms. The plate drilling program is huge and not designed for this task. We were afraid to touch it. But, by writing some SQL and playing with the data, we were able to develop a drilling plan in an evening. Over the ensuing 2 months the plates were drilled, used for observation, and the data was reduced. Within an hour of getting the data, they were loaded into the SkyServer database and we have used them to improve the redshift predictor — it became much more accurate on that class of galaxies. Now others are asking our help to design specialized plates for their projects.

We believe these two experiences and many similar ones, along with the 20+15 queries in the appendix, are a very promising sign that commercial database tools can indeed help scientists organize their data for data mining and easy access.

## Acknowledgements

We acknowledge our obvious debt to the people who built the SDSS telescope, those who operate it, those who built the SDSS processing pipelines, and those who operate the Fermilab pipeline. The SkyServer data depends on the efforts of all those people. In addition Robert Lupton has been very helpful in explaining the photo-object processing and some of the subtle meanings of the attributes, Mark Subbarao has been equally helpful in explaining the spectrogram attributes and Steve Kent has helped us to understand the observations better. James Annis, Xiaohui Fan, Gordon Richards, Michael Strauss, and Paula Szkody helped us compose some of the more complex queries. David DeWitt helped us improve the presentation. We thank Compaq and Microsoft for donating the project's hardware and software.

## Appendix: A Detailed Narrative of the Twenty Queries

This section presents each query, its translation to SQL, and a discussion of how the Query performs on the SkyServer at Fermi Lab. The computer is a Compaq ProLiant Ml530 with two 1GHz Pentium III Xeon processors, 2GB of 133MHz SDRAM; a 64bit/66MHz PCI bus with eight 10K rpm SCSI disks configured as 4 mirrored volumes. The database, log, and temporary database, and logs are all spread across these disks.

Some queries first define constants (see for example query 1) that are later used in the query – rather than calling the constant function within the query. If we do not do this, the SQL query optimizer takes the very conservative view that the function is not a constant and so the query plan calls the function for every tuple. It also suspects that the function may have side effects, so the optimizer turns off parallelism. So, function calls inside queries cause a 10x or more slowdown for the query and corresponding CPU cost increase. As a workaround, we rarely use functions within a query – rather we define variables (e.g. @saturated in Q1) and assign the function value to the variable before the query runs. Then the query uses these (constant) variables.

**Q1: Find all galaxies without saturated pixels within 1' of a given point.**

The query uses the table valued function `getNearbyObjEq()` that does an HTM cover search to find nearby objects. This handy function returns the object's ID, distance, and a few other attributes. The query also uses the *Galaxy* view to filter out everything but primary (good) galaxy objects.

```sql
declare @saturated bigint;                          -- initialized "saturated" flag
set    @saturated = dbo.fPhotoFlags('saturated');   -- avoids SQL2K optimizer problem
select G.objID, GN.distance                         -- return Galaxy Object ID and
into   ##results                                    -- angular distance (arc minutes)
from   Galaxy                       as G            -- join Galaxies with
 join  fGetNearbyObjEq(185,-0.5, 1) as GN           -- objects within 1' of ra=185 & dec=-.5
                on G.objID = GN.objID               -- connects G and GN
where  (G.flags & @saturated) = 0                   -- not saturated
order by distance                                   -- sorted nearest first
```

The query returns 19 galaxies in 50 milliseconds of CPU time and 0.19 seconds of elapsed time. The following picture shows the query plan (the rows from the table-valued function `GetNerabyObjEQ()` are nested-loop joined with the *photoObj* table – each row from the function is used to probe the photoObj table to test the *saturated* flag, the *primary object* flag, and the *galaxy* type.). The function returns 22 rows that are joined with the photoObj table on the ObjID primary key to get the object's flags. 19 of the objects are not saturated and are primary galaxies, so they are sorted by distance an inserted in the ##results temporary table.

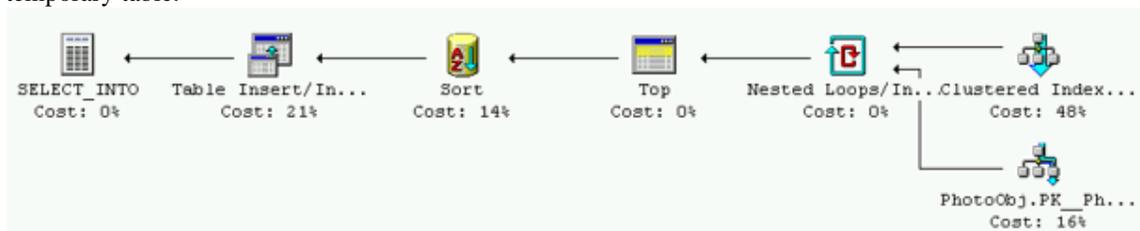

**Q2: Find all galaxies with blue surface brightness between and 23 and 25 magnitude per square arcseconds, and super galactic latitude (sgb) between (-10º, 10º), and declination less than zero.**

The surface brightness is defined as the logarithm of flux per unit area on the sky. Since the magnitude is -2.5 log(flux), the SB is $-2.5 \log(flux/R^2 \pi)$. The SkyServer pipeline precomputed the value rho = -5 log( R ) – 2.5 log ($\pi$), where R is the radius of the galaxy. Thus, for a constraint on the surface brightness in the *g* band we can use the combination *g+rho*.

```sql
select objID                  -- Get the object identifier
into ##results
from Galaxy                   -- of all the galaxies that have
where  ra between 170 and 190 -- designated ra/dec   (need galactic coordinates)
 and dec < 0                  -- declination less than zero.
 and g+rho between 23 and 25  -- g = blue magnitude,
                              -- rho= 5*ln(r)
                              -- g+rho = SB per sq arc sec is between 23 and 25
```

This query finds 191,062 objects in 18.6 seconds elapsed, 14 seconds of CPU time. This is a parallel scan of the *XYZ* index of the *PhotoObj* table (*Galaxy* is a view of that table that only shows primary objects that are of type Galaxy). The XYZ index covers this query (contains all the necessary fields). The query spends 2 seconds inserting the answers in the *##results* set, if the query just counts the objects, it runs in 16 seconds.

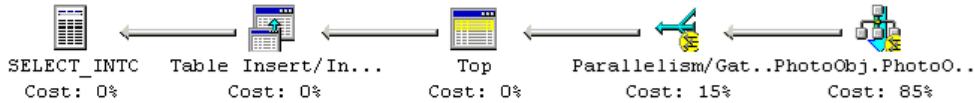

**Q3: Find all galaxies brighter than magnitude 22, where the local extinction is >0.175.**
The extinction indicates how much light is absorbed by that dust that is between the object and the earth. There is an extinction table, giving the extinction for every "cell", but the extinction is also stored as an attribute of each element of the *PhotoObj* table, so the simple query is:

```
select  objID                          -- find the object IDs
 into ##results
 from Galaxy                           -- join Galaxies with Extinction table
    where       r < 22                 -- where brighter than 22 magnitude
    and         reddening_r> 0.175     -- extinction more than 0.175
```

The query returns 488,183 objects in 168 seconds and 512 seconds of CPU time – the large CPU time reflects an SQL feature affectionately known as the "bookmark bug". SQL thinks that very few galaxies have r<22, so it finds those in the index and then looks up each one to see if it has reddening_r > .175. We could force it to just scan the base table (by giving it a hint), but that would be cheating. The query plan does a sequential scan of the 14 million records in the *PhotoObj.xyz* index to find the approximately 500,000 galaxy objIDs that have magnitude less than 22. Then it does a lookup of each of these objects in the base table (1/2 a million "bookmark" lookups) to check the reddening. The query uses about 30% of one of the two CPUs – much of this is spent inserting the ½ million answer records. If the extinction matrix were used, this query could use the HTM index and run about five times faster. The choice of a bookmark lookup may be controversial, but it does run quickly.

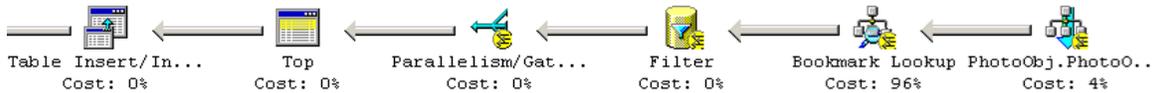

**Q4: Find galaxies with an isophotal surface brightness (SB) larger than 24 in the red band, with an ellipticity>0.5, and with the major axis of the ellipse between 30" and 60" arc seconds (a large galaxy).**
Each of the five color bands has been pre-processed into a bitmap image that is broken into 15 concentric rings. The rings are further divided into octants. This information is stored in the object's *profile*. The intensity of the light in each ring and octant is pre-processed to compute surface brightness, ellipticity, major axis, and other attributes. These derived attributes are stored with the *PhotoObj*, so the query operates on these derived quantities.

```
select ObjID                    -- put the qualifying galaxies in a table
into ##results
from Galaxy                     -- select galaxies
where r + rho < 24              -- brighter than magnitude 24 in the red spectral band
  and  isoA_r between 30 and 60 -- major axis between 30" and 60"
  and  (power(q_r,2) + power(u_r,2)) > 0.25 -- square of ellipticity is > 0.5 squared.
```

The query returns 787 rows in 18 seconds elapsed, 9 seconds of CPU time. It does a parallel scan of the NEO index on the photoObj Table that covers the object *type*, *status*, *flags*, and also *isoA*, *q_r,* and *r*. The query then does a bookmark lookup on the qualifying galaxies to check the $r+rho$ and $q\_r^2 + u\_r^2$ terms. The resulting records are inserted in the answer set.

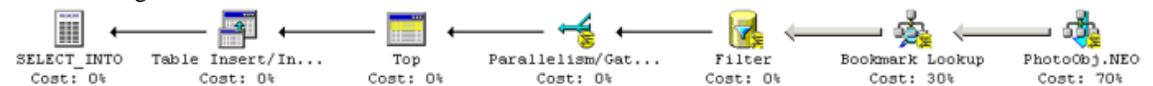

**Q5: Find all galaxies with a deVaucouleours profile (r¼ falloff of intensity on disk) and the photometric colors consistent with an elliptical galaxy.** As discussed in Q4, the deVaucouleours profile information is precomputed from the concentric rings during the pipeline processing. There is a likelihood value stored in the table, which tells whether the deVaucouleours profile or an exponential disk is a better fit to the galaxy.

```
declare  @binned        bigint;                              -- initialized "binned" literal
set      @binned =      dbo.fPhotoFlags('BINNED1') +         -- avoids SQL2K optimizer problem
                        dbo.fPhotoFlags('BINNED2') +
                        dbo.fPhotoFlags('BINNED4') ;
declare  @blended       bigint;                              -- initialized "blended" literal
set      @blended =     dbo.fPhotoFlags('BLENDED');          -- avoids SQL2K optimizer problem
declare  @noDeBlend     bigint;                              -- initialized "noDeBlend" literal
set      @noDeBlend =   dbo.fPhotoFlags('NODEBLEND');        -- avoids SQL2K optimizer problem
declare  @child         bigint;                              -- initialized "child" literal
set      @child =       dbo.fPhotoFlags('CHILD');            -- avoids SQL2K optimizer problem
declare  @edge          bigint;                              -- initialized "edge" literal
set      @edge =        dbo.fPhotoFlags('EDGE');             -- avoids SQL2K optimizer problem
declare  @saturated     bigint;                              -- initialized "saturated" literal
set      @saturated =   dbo.fPhotoFlags('SATURATED');        -- avoids SQL2K optimizer problem
select objID
into ##results
from Galaxy as G              -- count galaxies
where  lDev_r > 1.1 * lExp_r  -- red DeVaucouleurs fit likelihood greater than disk fit
   and lExp_r > 0             -- exponential disk fit likelihood in red band > 0
   -- Color cut for an elliptical galaxy courtesy of James Annis of Fermilab
   and (G.flags & @binned) > 0
   and (G.flags & ( @blended + @noDeBlend + @child)) != @blended
   and (G.flags & (@edge + @saturated)) = 0
   and (G.petroMag_i > 17.5)
   and (G.petroMag_r > 15.5 OR G.petroR50_r > 2)
   and (G.petroMag_r < 30 and G.g < 30 and G.r < 30 and G.i < 30)
   and ((G.petroMag_r-G.reddening_r) < 19.2)
   and (    (       ((G.petroMag_r - G.reddening_r) < (13.1 +   -- deRed_r < 13.1 +
                                        (7/3)*(G.g - G.r) +     -- 0.7 / 0.3 * deRed_gr
                                  4 *(G.r - G.i) -4 * 0.18 ))   -- 1.2 / 0.3 * deRed_ri
              and (( G.r - G.i - (G.g - G.r)/4 - 0.18) BETWEEN -0.2 AND  0.2 )
             )
         or
           (     (( G.petroMag_r - G.reddening_r) < 19.5 )      -- deRed_r < 19.5 +
             and (( G.r - G.i -(G.g - G.r)/4 -.18) >            -- cperp = deRed_ri
                       (0.45 - 4*( G.g - G.r)))                 -- 0.45 - deRed_gr/0.25
             and ((G.g - G.r) > ( 1.35 + 0.25 *(G.r - G.i)))
           )    )
```

The query found 40,005 objects in 166 seconds elapsed, 66 seconds of CPU time. This is parallel table scan of *PhotoObj* table because there is no covering index. The fairly complex query evaluation all hides in the parallel scan and parallel filter nodes at the right of the figure below.

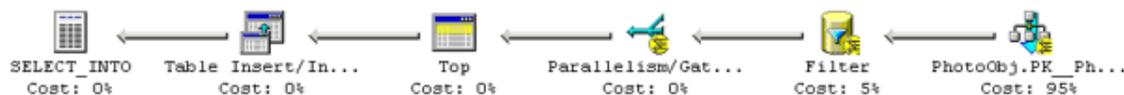

**Q6: Find galaxies that are blended with a star and output the deblended galaxy magnitudes.**

Some objects overlap others. The most common cases are a star in front of a galaxy or a star in the halo of another star. These "deblended" objects, record their "parent" objects in the database. So this query starts with a deblended galaxy (one with a parent) and then looks for all stars that have the same parent. It then outputs the five color magnitudes of the star and the parent galaxy.

```
select G.ObjID, G.u, G.g, G.r, G.i, G.z            -- output galaxy and magnitudes.
into   ##results
from   galaxy G, star S                            -- for each galaxy
where  G.parentID > 0                              -- galaxy has a "parent"
  and  G.parentID = S.parentID                     -- star has the same parent
```

The query found 1,088,806 galaxy-star pairs in 41 seconds. Without an index on the parent attribute, this is a Cartesian product of two very large tables and would involve about $10^{16}$ join steps. So, it makes good sense to create an index or intermediate table that has the deblended stars. Fortunately, SkyServer already has a parent index on the *photoObj* table, since we often want to find the children of a common parent. The clause *parentID>0* excludes galaxies with no parent. These two steps cut the task from about $10^{20}$ down to a near-linear $10^8$ steps (because ½ the objects are galaxies and about 25% of them have parents). The plan scans the Parent index and builds a hash table of *parent IDs*, *galaxy IDs* that have parents (about 3.7M objects, so about 40MB). It then scans over the index a second time looking at stars that have parents. It looks in the hash table to see if the parent is also a parent of a galaxy. If so, the *galaxy ID* and *star ID* are inserted in the answer set.

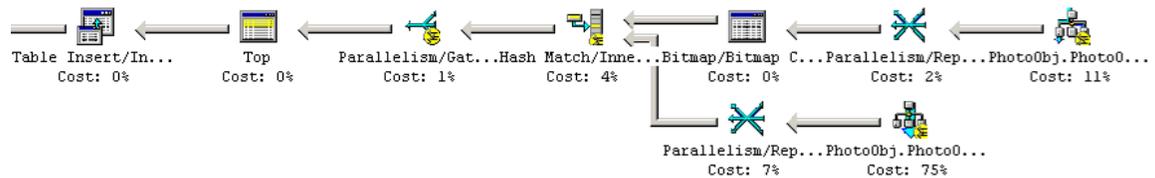

**Q7: Provide a list of star-like objects that are 1% rare.**

The survey gets magnitude information about stars in 5 color bands. This query looks at the ratios of the brightness in each band. (Luminosity ratios are magnitude differences because magnitudes are logarithms of the actual brightness in that band). The query "bins" these magnitudes based on the 4-space of u-g, g-r, r-i, i-z. Experimentation showed that dividing the bins in integer units worked well. We built a results table that contains all the bins. The large-population bins are deleted, leaving only the rare ones (less than 500 members).

```
select  cast(round((u-g),0) as int) as UG,
        cast(round((g-r),0) as int) as GR,
        cast(round((r-i),0) as int) as RI,
        cast(round((i-z),0) as int) as IZ,
        count(*)                    as pop
into   ##results
from   star
where  (u+g+r+i+z) < 150   -- exclude bogus magnitudes (== 999)
group by       cast(round((u-g),0) as int), cast(round((g-r),0) as int),
               cast(round((r-i),0) as int), cast(round((i-z),0) as int)
order by count(*)
```

This query found 15,528 buckets in less than a minute. The first 140 buckets have 99% of the objects. The query scans the *UGRIZ* index of the *photoObj* table in parallel to populate a hash table containing the counts. When the scan is done, the hash table is sorted put into the results table. The query uses 90 seconds of CPU time in 53 seconds elapsed time (this is a dual processor system).

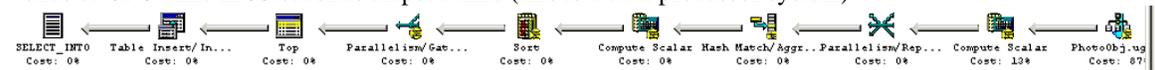

OK, now use this as a filter to return rare stars.
```
delete   ##results
where pop > 500
```
This whole scenario uses less than 2 minutes of computer time.

**Q8: Find all objects with unclassified spectra.**

A search for all objects that have spectra that do not match any known category.

```
declare @unknown bigint;                           -- initialized "binned" literal
set    @unknown = dbo.fSpecClass('UNKNOWN')
select specObjID
into   ##results
from   SpecObj
where  SpecClass = @unknown
```

This is a simple scan of the *SpectraObj* table looking for those spectra that have not yet been classified. It finds 260 rows in .126 seconds and .03 seconds of CPU time.

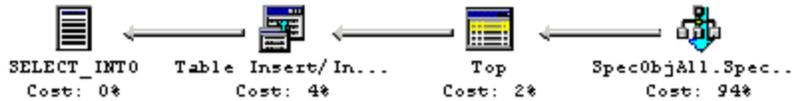

**Q9: Find quasars with a line width >2000 km/s and 2.5<redshift<2.7.**

This is a sequential scan of quasars in the Spectra table with a predicate on the redshift and line width. The Spectra table has about 53 thousand objects having a known spectrum but there are only 4,300 known quasars. We need to do a join with the SpecLine table, for all the emission lines used in the redshift determination, look for the highest amplitude one. The line width can be computed from the sigma attribute of the line, which is the width of the line in Angstroms. The conversion to km/s is : *lineWidth = sigma * 300000 / wave*.

```
declare      @qso     int;
set          @qso = dbo.fSpecClass('QSO') ;
declare      @hiZ_qso int;
set          @hiZ_qso =dbo.fSpecClass('HIZ-QSO');
select       s.specObjID,                          -- object id
             max(l.sigma*300000.0/l.wave) as veldisp,  -- velocity dispersion
             avg(s.z) as z                         -- redshift
into   ##results
from     SpecObj s, specLine l                     -- from the spectrum table and lines
where  s.specObjID=l.specObjID                     -- line belongs to spectrum of this obj
   and ( (s.specClass = @qso) or                   -- quasar
         (s.specClass = @hiZ_qso))                 -- or hiZ_qso.
   and  s.z between 2.5 and 2.7                    -- redshift of 2.5 to 2.7
   and  l.sigma*300000.0/l.wave >2000.0            -- convert sigma to km/s
   and  s.zConf > 0.9                              -- high confidence on redshift estimate
group by s.specObjID
```

This is a sequential scan of the Spectra table with a predicate looking for quasars with the specified redshift (and good credibility on the redshift estimate). When it finds such a quasar, it does a nested loops join with the spectral lines to see if they have acceptable line width. The Spectra table has about 53 thousand objects having a known spectrum but there are only 4,300 known quasars. The acceptable spectra (and their lines are passed to an aggregator that computes the maximum velocity and the average redshift. The query returns 54 rows in 436 ms.

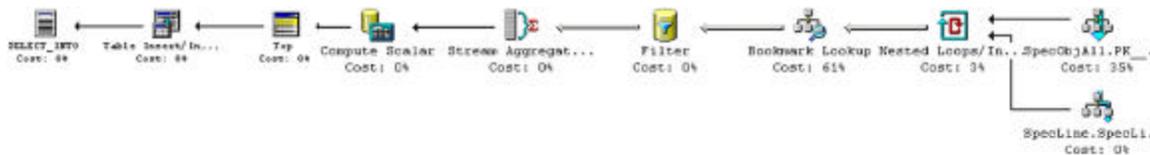

**Q10: Find galaxies with spectra that have an equivalent width in Ha >40Å (Ha is the main hydrogen spectral line.)**

This is a simple 4-way join of Galaxies with Spectra and then their lines and then the line names.

```
select G.ObjID                        -- return qualifying galaxies
 into   ##results
 from   Galaxy       as G,            -- G is the galaxy
        SpecObj      as S,            -- S is the spectra of galaxy G
        SpecLine     as L,            -- L is a line of S
        specLineNames as LN           -- the names of the lines
 where  G.ObjID = S.ObjID             -- connect the galaxy to the spectrum
   and S.SpecObjID = L.SpecObjID      -- L is a line of S.
   and L.LineId = LN.value            -- L is the H alpha line
   and LN.name =   'Ha_6565'
   and L.ew > 40                      -- H alpha is at least 40 angstroms wide.
```

This query runs in parallel and uses 5 CPU seconds in 5 seconds of elapsed time. It finds 5,496 galaxies with the desired property. Interestingly, SQL decides to do this query inside-out. It first finds all lines that qualify, then it finds the parent spectra, and then it sees if the parent spectrum is a galaxy. The middle join is a parallel hash join; while the inner and outer are nested loops joins (qualifying spectra with photo objects).

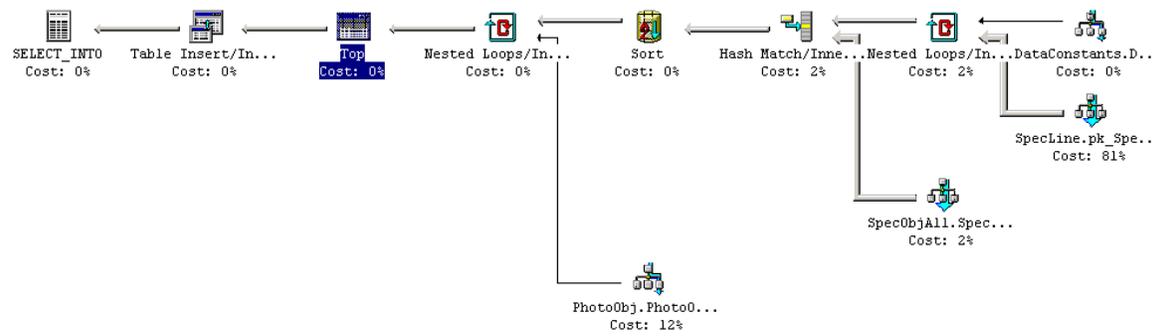

That was easy, so lets also find objects with a weak Hbeta line (Halpha/Hbeta > 20.)

```
select G.ObjID                          -- return qualifying galaxies
 into   ##results
 from   Galaxy       as G,              -- G is the galaxy
        SpecObj      as S,              -- S is the spectra of galaxy G
        SpecLine     as L1,             -- L1 is a line of S
        SpecLine     as L2,             -- L2 is a second line of S
        specLineNames as LN1,           -- the names of the lines (Halpha)
        specLineNames as LN2            -- the names of the lines (Hbeta)
where  G.ObjID = S.ObjID                -- connect the galaxy to the spectrum
  and S.SpecObjID = L1.SpecObjID        -- L1 is a line of S.
  and S.SpecObjID = L2.SpecObjID        -- L2 is a line of S.  and L1.LineId = LN1.LineId
  and L1.LineId = LN1. value
  and LN1.name =   'Ha_6565'                 -- L1 is the H alpha line
  and L2.LineId = LN2.value             -- L2 is the H alpha line
  and LN2.name =   'Hb_4863'            --
  and L1.ew > 200                       -- BIG Halpha
  and L2.ew > 10                        -- significant Hbeta emission line
  and L2.ew * 20 < L1.ew                -- Hbeta is comparatively small
```

This query uses 1.9 seconds of CPU time in 1.3 seconds elapsed time to return 9 objects. It is slightly more complex than the plan for query 10, involving two more nested loops joins.

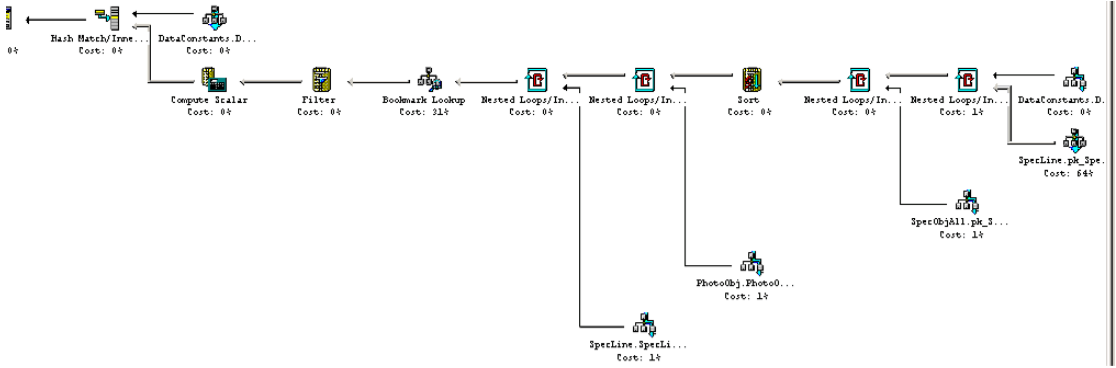

**Q11: Find all elliptical galaxies with spectra that have an anomalous emission line.**

This is a search for galaxies that match the elliptical template, and that have an "unknown" spectral line with the property that there is no nearby (within 0.01 angstroms) line that has been identified.

```sql
select  distinct G.ObjID              -- return qualifying galaxies
into    ##results
from    Galaxy       as G,            -- G is the galaxy
        SpecObj      as S,            -- S is the spectra of galaxy G
        SpecLine     as L,            -- L is a line of S
        specLineNames as LN,          -- the type of line
        XCRedshift   as XC            -- the template cross-correlation
where G.ObjID = S.ObjID               -- connect galaxy to the spectrum
  and S.SpecObjID = L.SpecObjID       -- L is a line of S
  and S.SpecObjID = XC.SpecObjID      -- CC is a cross-correlation with templates
  and XC.tempNo = 8                   -- Template('Elliptical') -- CC says "elliptical"
  and L.LineID = LN.value             -- line type is found
  and LN.Name = 'UNKNOWN'             --      but not identified
  and L.ew > 10                       -- a prominent (wide) line
  and S.SpecObjID not in (            -- insist that there are no other lines
      select S.SpecObjID              -- that are know and are very close to this one
      from   SpecLine      as L1,     -- L1 is another line
             specLineNames as LN1
      where S.SpecObjID = L1.SpecObjID     -- for this object
        and abs(L.wave - L1.wave) <.01     -- at nearly the same wavelength
        and L1.LineID = LN1.value          -- line found and
        and LN1.Name != 'UNKNOWN'          --      it IS identified
      )
```

This query finds 22 thousand galaxies in 33 seconds of CPU time and 28 seconds of clock time. It starts by building a list of all the prominent unknown lines (the nested loops join at the far right of the picture below). Then it joins that list with the SpecObj table. Then it does a nested loops join with photoObj to discard objects that are not galaxies. Now it does a nested loops join of the lines of that specObj that are nearby the unknown line and are in fact know. If none are fond, then it does a hash join with the *XCredshift* table to discard any galaxy that does not qualify as elliptical.

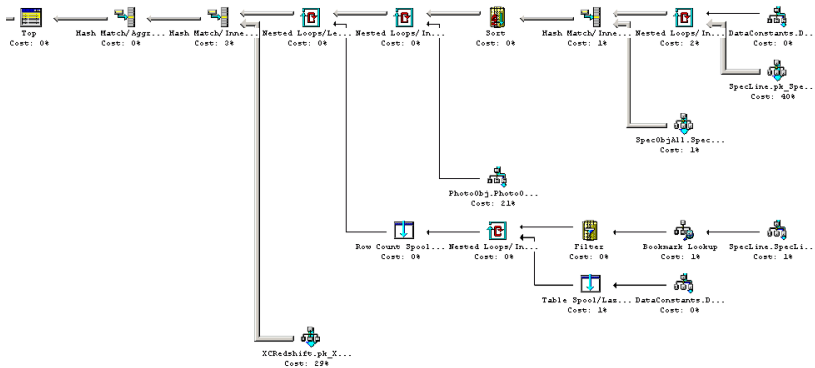

## Q12: Create a grided count of galaxies with u-g>1 and r<21.5 over -5<declination<5, and 175<right ascension<185, on a grid of 2' arc minutes. Create a map of masks over the same grid.

Scan the table for galaxies and group them in cells 2 arc-minutes on a side filtering with predicates on the u-g magnitude ratio and the r magnitude. To limit the search to the portion of the sky defined by the right ascension and declination conditions, the query uses the `fHTM_Cover()` procedure to constrain the HTM ranges of candidate objects. The query returns the count of qualifying galaxies in each cell – 26,669 cells in all. We then run a second query with the same grouping, but with a predicate to include only objects such as satellites, planets, and airplanes that obscure the cell. The second query returns a list of cell coordinates that serve as a mask for the first query – 135 cells in all. The mask is stored in a temporary table and may be joined with the first query to delete or mask cells erroneous cells.

```sql
--- First find the grided galaxy count (with the color cut)
--- In local tangent plane, ra/cos(dec) is a "linear" degree.
declare @LeftShift16 bigint;           -- used to convert 20-deep htmIds to 6-deep IDs
set     @LeftShift16 = power(2,28);
select cast((ra/cos(cast(dec*30 as int)/30.0))*30 as int)/30.0 as raCosDec,
       cast(dec*30 as int)/30.0                               as dec,
       count(*)                                               as pop
into ##GalaxyGrid
from   Galaxy as G ,
       dbo.fHTM_Cover('CONVEX J2000 6 6 175 -5 175 5 185 5 185 -5') as T
where  htmID between T.HTMIDstart*@LeftShift16 and T. HTMIDend*@LeftShift16
  and  ra between 175 and 185
  and  dec between -5 and 5
  and  u-g > 1
  and  r < 21.5
group by  cast((ra/cos(cast(dec*30 as int)/30.0))*30 as int)/30.0,
          cast(dec*30 as int)/30.0
```

This query first builds a 6-deep htm mesh and then does a nested-loops join on the HTMindex of the *PhotoObj* table. The matching htm tuples are checked for acceptable ra and dec, ane u-g, and r values. If they pass this test they are streamed to a hash aggregation table and added to the appropriate bin. When the scan is complete, the counts in the hash aggregate table is dumped to the answer set.

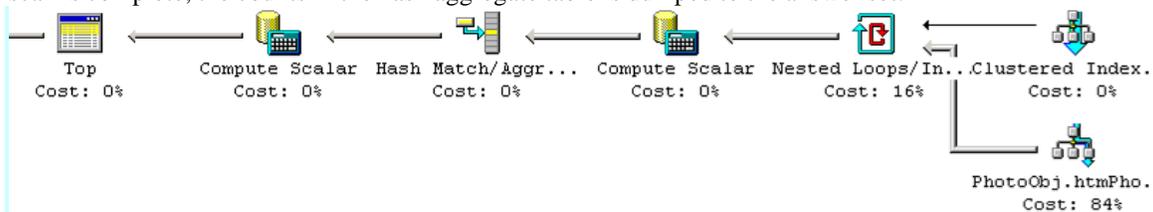

```sql
--- now build mask grid.
select cast((ra/cos(cast(dec*30 as int)/30.0))*30 as int)/30.0 as raCosDec,
       cast(dec*30 as int)/30.0                               as dec,
       count(*)                                               as pop
into ##MaskGrid
from   photoObj as PO,
       dbo.fHTM_Cover('CONVEX J2000 6 6 175 -5 175 5 185 5 185 -5') as T,
       photoType as PT
where  htmID between T.HTMIDstart*@LeftShift16 and T. HTMIDend*@LeftShift16
  and ra between 175 and 185
  and dec between -5 and 5
  and PO.type = PT.value
  and PT.name in ('COSMIC_RAY', 'DEFECT', 'GHOST', 'TRAIL', 'UNKNOWN')
group by  cast((ra/cos(cast(dec*30 as int)/30.0))*30 as int)/30.0,
          cast(dec*30 as int)/30.0
```

This query is similar to the previous one except that the photo objects are filtered by the bad flags.

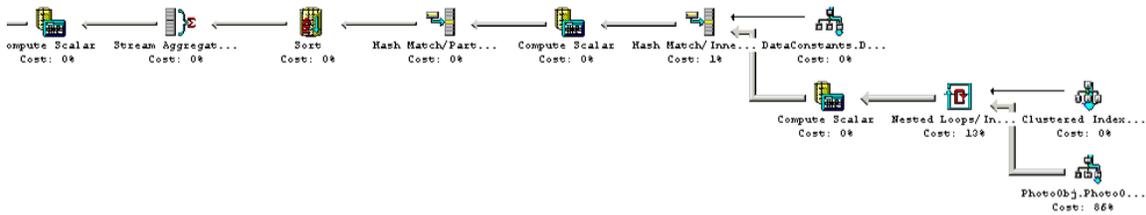

**Q13: Create a count of galaxies for each of the HTM triangles which satisfy a certain color cut, like 0.7u-0.5g-0.2i<1.25 and r<21.75, output it in a form adequate for visualization.**

Return the ra and dec and count for every 8-deep htm bucket (each bucket is about 1 square degree).

```
declare @RightShift12 bigint;
set     @RightShift12 = power(2,24);
select (htmID /@RightShift12) as htm_8,  -- group by 8-deep HTMID (rshift HTM by 12)
       avg(ra)   as ra,
       avg(dec)  as [dec],
       count(*)  as pop                -- return center point and count for display
 into   ##results                      -- put the answer in the results set.
 from   Galaxy                         -- only look at galaxies
 where  (0.7*u - 0.5*g - 0.2*i) < 1.25   -- meeting this color cut
   and  r < 21.75                      -- fainter than 21.75 magnitude in red band.
 group by (htmID /@RightShift12)       -- group into 8-deep HTM buckets..HTM buckets
```

The query returns 7,604 buckets in 20 seconds. A hash aggregate based on a scan of the xyz index, runs in 20 seconds elapsed, 28 seconds of CPU time.

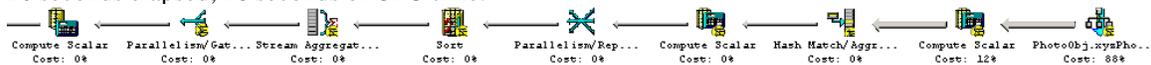

**Q14: Find stars with multiple measurements that have magnitude variations >0.1.**

This is a spatial join of the PhotoObj table with itself using the neighbors table to cut down on the search. Stars within ½ arcsecond of one another in two different observations are considered to be identical (½ second is approximately the minimum telescope resolution.)

```
declare @star int;                          -- initialized "star" literal
set     @star = dbo.fPhotoType('Star');     -- avoids SQL2K optimizer problem
select s1.objID as ObjID1, s2.objID as ObjID2 -- select object IDs of star and its pair
into   ##results
from   star      as  s1,                    -- the primary star
       photoObj  as  s2,                    -- the second observation of the star
       neighbors as  N                      -- the neighbor record
where s1.objID = N.objID                    -- insist the stars are neighbors
  and s2.objID = N.neighborObjID            -- using precomputed neighbors table
  and distanceMins < 0.5/60                 -- distance is ½ arc second or less
  and s1.run != s2.run                      -- observations are two different runs
  and s2.type = @star                       -- s2 is indeed a star
  and  s1.u between 1 and 27                -- S1 magnitudes are reasonable
  and  s1.g between 1 and 27
  and  s1.r between 1 and 27
  and  s1.i between 1 and 27
  and  s1.z between 1 and 27
  and  s2.u between 1 and 27                -- S2 magnitudes are reasonable.
  and  s2.g between 1 and 27
  and  s2.r between 1 and 27
  and  s2.i between 1 and 27
  and  s2.z between 1 and 27
  and (                                     -- and one of the colors is  different.
         abs(S1.u-S2.u) > .1 + (abs(S1.Err_u) + abs(S2.Err_u))
      or abs(S1.g-S2.g) > .1 + (abs(S1.Err_g) + abs(S2.Err_g))
      or abs(S1.r-S2.r) > .1 + (abs(S1.Err_r) + abs(S2.Err_r))
      or abs(S1.i-S2.i) > .1 + (abs(S1.Err_i) + abs(S2.Err_i))
      or abs(S1.z-S2.z) > .1 + (abs(S1.Err_z) + abs(S2.Err_z))
      )
```

This is a parallel merge join of the *neighbors* table with the PhotoObj table to find all stars that are within ½ arcsecond of some other object and such that the star has a reasonable magnitude.  The result of that join is parallel hash match join with the PhotoObj table filtered by the "reasonable magnitude" predicate,  that join feeds to a filter that discards objects where the differences of the magnitudes is less than the threshold. The query runs in 118 seconds and uses 108 CPU seconds to find 48,245 such stars.

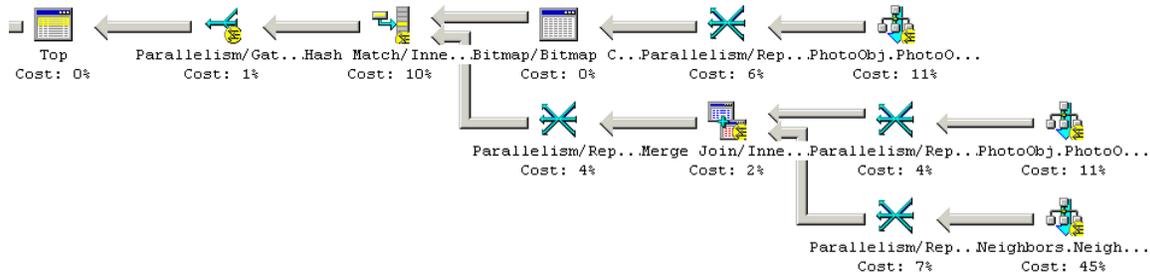

**Q15: Provide a list of moving objects consistent with an asteroid.**

Objects are classified as moving if their positions change over the time of observation. SDDS makes 5 successive observations from the 5 color bands over a 5 minute period. If an object is moving, the successive images see a moving image against the fixed background of the galaxies. The processing pipeline computes this movement velocity as rowV (the row velocity) and colV the column velocity. So query 15 becomes a simple table scan computing the velocities and selecting those objects that have high velocity.

```
select objID,                                         -- return object ID
       sqrt( power(rowv,2) + power(colv, 2) ) as velocity, -- velocity
       dbo.fGetUrlExpId(objID) as Url                 -- url of image to examine it.
into   ##results
from   PhotoObj                                       -- check each object.
where (power(rowv,2) + power(colv, 2)) between 50 and 1000 -- square of velocity
  and rowv >= 0 and colv >=0                          -- negative values indicate error
```

This is a sequential scan of the PhotoObj table (there is no covering index. It uses 72 seconds of CPU time in 162 second of elapsed time to evaluate the predicate on each of the 14M objects. It finds 1,303 candidates. Here is a picture of one of the objects: 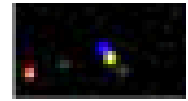

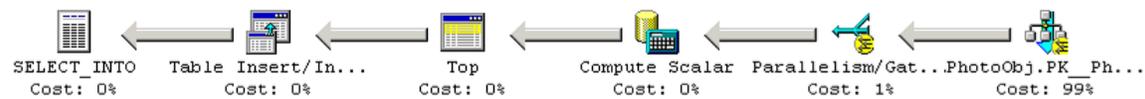

These are "slow moving" objects. To find fast moving objects we write a different query (based on a tcl script written by Steve Kent). This query looks for streaks in the sky that line up. These streaks are not close enough to be identified as a single object. The query starts out with all pairs of objects in a given area (run,camcol,field) that have a fiberMag_r between 6 and 22, and are elongated. We first select the red and green candidates, by requiring that they are fainter in all the other colors. These two are much rarer than the i candidates. Also, there is one candidate, where the i' image is blended with the r'. Next we do a join on these two, and also require that the magnitudes in g and r are within 2, and they are within 4 arcminutes of one another, in the same run and camcol, but they can be on adjacent fields. We found 4 pairs, in one of them the red objects is degenerate, probably deblended. Each of the other three is a NEO.

```
select  r.objID as rId, g.objId as gId,
        r.run, r.camcol,
        r.field as field, g.field as gField,
        r.ra as ra_r, r.dec as dec_r,
        g.ra as ra_g, g.dec as dec_g,  --(note acos(x) ~ x for x~1)
        sqrt(power(r.cx-g.cx,2)+power(r.cy-g.cy,2)+power(r.cz-g.cz,2)) *
             (180*60/PI()) as distance,
        dbo. fGetUrlExpId (r.objID) as rURL,    -- returns URL for image of object
        dbo. fGetUrlExpId (g.objID) as gURL
from    PhotoObj r, PhotoObj g
where   r.run = g.run and r.camcol=g.camcol   -- same run and camera column
  and abs(g.field-r.field) <= 1               -- adjacent fields
       -- the red selection criteria
  and ((power(r.q_r,2) + power(r.u_r,2)) > 0.111111 )  -- q/u is ellipticity
  and r.fiberMag_r between 6 and 22
  and r.fiberMag_r < r.fiberMag_u
  and r.fiberMag_r < r.fiberMag_g
  and r.fiberMag_r < r.fiberMag_i
  and r.fiberMag_r < r.fiberMag_z
  and r.parentID=0
  and r.isoA_r/r.isoB_r > 1.5
  and r.isoA_r > 2.0
       -- the green selection criteria
  and ((power(g.q_g,2) + power(g.u_g,2)) > 0.111111 )
  and g.fiberMag_g between 6 and 22
  and g.fiberMag_g < g.fiberMag_u
  and g.fiberMag_g < g.fiberMag_r
  and g.fiberMag_g < g.fiberMag_i
  and g.fiberMag_g < g.fiberMag_z
  and g.parentID=0
  and g.isoA_g/g.isoB_g > 1.5
  and g.isoA_g > 2.0
-- the match-up of the pair  --(note acos(x) ~ x for x~1)
  and sqrt(power(r.cx-g.cx,2)+power(r.cy-g.cy,2)+power(r.cz-g.cz,2))*(180*60/pi()) < 4.0
  and abs(r.fiberMag_r-g.fiberMag_g)< 2.0
```

This query is a scan of the NEO index that that has the fiberMag array and also the iso parameters.  It is nested loops join of this array with itself on the run, camcol, field keys, doing a nested loops join, for each object that qualifies in the red band, finding all the qualifying green objects.  When it finds a matching pair, it checks to see if the parentID is 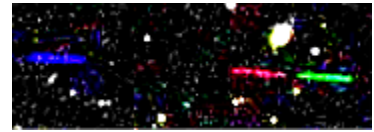
zero for both of them.  It with the index it finds 4 objects in 55 seconds elapsed and 51 seconds of CPU time.  Without the NEO index it takes about 10 minutes.

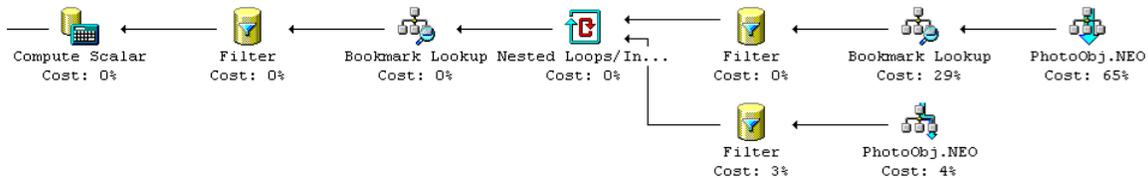

You can see one of the example objects at:
http://skyServer.sdss.org/en/tools/navi/getimg.asp?zoom=1&ra=171.161254&dec=-0.0108500

### Q16: Find all objects similar to the colors of a quasar at 5.5<redshift<6.5.

Scan all objects with a predicate to select objects that satisfy the quasar color cut.  Produce a table counting the total, the objects classed as galaxies, those classed as stars and those classed as "other".  This is a parallel sequential scan of the ugriz index. It runs in 15 seconds of CPU time (1 M records per second) and 14 seconds of clock time. It is IO bound, needing 15 seconds to read the index at 80MBps. It finds 1,826 objects, 1,489 Galaxies, and 337 stars.

```
select count(*)                                          as 'total',
sum( case when (type=3) then 1 else 0 end)               as 'Galaxies',
sum( case when (type=6) then 1 else 0 end)               as 'Stars',
sum( case when (type not in (3,6)) then 1 else 0 end) as 'Other'
from    PhotoPrimary                          -- for each object
 where (( u - g > 2.0) or (u > 22.3) )        -- apply the quasar color cut.
   and ( i between 0 and 19 )
   and ( g - r > 1.0 )
   and ( (r - i < 0.08 + 0.42 * (g - r - 0.96)) or (g - r > 2.26 ) )
   and ( i - z < 0.25 )
```

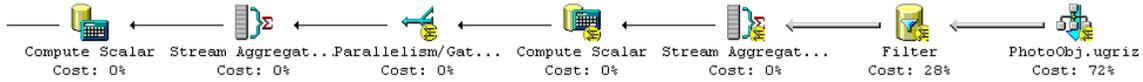

### Q17: Find binary stars where at least one of them has the colors of a white dwarf.

White dwarf color cut courtesy of Paul Szkody.

```
declare @star int;                              -- initialized "star" literal
set     @star = dbo.fPhotoType('Star');         -- avoids SQL2K optimizer problem
select  s1.objID as s1, s2.objID as s2          -- return star pairs
into    ##results
from    Star       S1,                          -- S1 is the white dwarf
        Neighbors  N,                           -- N is the precomputed neighbors links
        Star       S2                           -- S2 is the second star
  where S1.objID = N. objID                     -- S1 and S2 are neighbors-within 30 arc sec
    and S2.objID = N.NeighborObjID
    and N.NeighborObjType = @star               -- and S2 is a star
    and N.DistanceMins < .05                    -- the 3 arcsecond test
    and (S1.u - S1.g) < 0.4                     -- and S1 meets Paul Szkody's color cut for
    and (S1.g - S1.r) < 0.7                     -- white dwarfs.
    and (S1.r - S1.i) > 0.4
    and (S1.i - S1.z) > 0.4
```

The query finds 2,773 objects in 18 seconds. It scans the ugriz index of the *photoObj* table for stars with white dwarf colors. Then it does a nested-loops join with the neighbors table to find objects within 3 arcseconds of qualifying stars. Now it joins those objects with the photoObj table to make sure that the neighbor is a star.

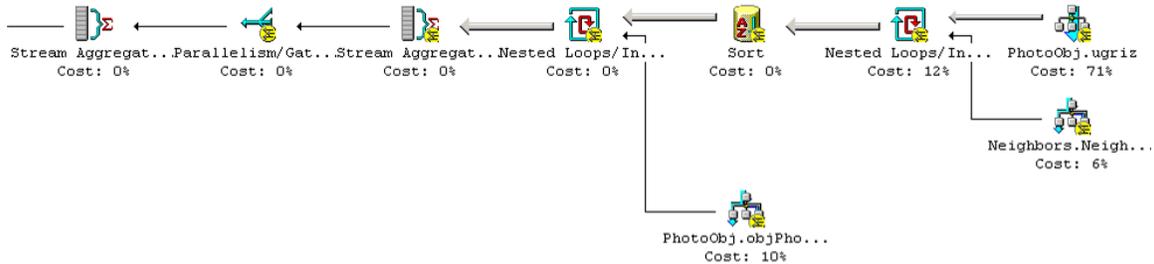

**Q18: Find all objects within 30 arcseconds of one another that have very similar colors: that is where the color ratios u-g, g-r, r-i are less than 0.05m.**

Magnitudes are logarithms so these differences are really ratios. This is a gravitational lens query. It scans for objects in the PhotoPrimary view and compares them to all objects within one arcminute of the object. If the color ratios match, this is a candidate object.

```sql
select distinct P.ObjID                    -- count distinct cases (will get min objid)
  into ##results                           -- oid compare gets minimum object
  From photoPrimary   P,                   -- P is the primary object
       Neighbors      N,                   -- N is the neighbor link
       photoPrimary   L                    -- L is the lens candidate of P
 where P.ObjID = N.ObjID                   -- N is a neighbor record
   and L.ObjID = N.NeighborObjID           -- L is a neighbor of P
   and P.ObjID < L.ObjID                   -- avoid duplicates
   and abs((P.u-P.g)-(L.u-L.g))<0.05       -- L and P have similar spectra.
   and abs((P.g-P.r)-(L.g-L.r))<0.05
   and abs((P.r-P.i)-(L.r-L.i))<0.05
   and abs((P.i-P.z)-(L.i-L.z))<0.05
```

This query finds 4,442 objects in 930 seconds elapsed and 1475 seconds of CPU time. It is the longest-running of the 20 queries. This is quite similar to Query14, it uses the neighbors table to join objects to nearby neighbors. It is a scan of ugriz index picking up the object's colors, and then a nested-loops join to the neighbors table which in turn is a join back to the PhotoObj ugriz. This query runs for a very long time because it is comparing every primary object to all its' neighbors. This is about 150 million comparisons (whereas Query 14 was comparing a few objects to their closest neighbor

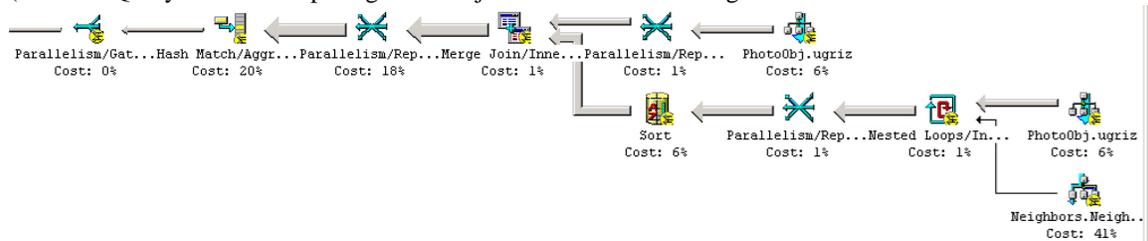

**Q19: Find quasars with a broad absorption line in their spectra and at least one galaxy within 10 arcseconds. Return both the quasars and the galaxies.**

Scan for quasars with a predicate for a broad absorption line and use them in a spatial join with galaxies that are within 10 arc-seconds. The neighbor's within 10 arcseconds is precomputed in the *neighbors* table, so join with that and subset the search.

```sql
select Q.ObjID as Quasar_candidate_ID, G.ObjID as Galaxy_ID
into ##results
from  SpecObj         as Q,           -- Q is the specObj of the quasar candidate
      Neighbors       as N,           -- N is the Neighbors list of Q
      Galaxy          as G,           -- G is the nearby galaxy
      SpecClass       as SC,
      SpecLine        as L,           -- L is the broad line we are looking for
      SpecLineNames   as LN
where Q.SpecClass  = SC.class
  and SC.name   in ('QSO', 'HIZ_QSO') -- Spectrum says "QSO"
  and Q.SpecObjID = L.SpecObjID       -- L is a spectral line of Q.
  and L.LineID = LN.value             -- line found and
  and LN.Name != 'UNKNOWN'            --    not not identified
  and L.ew < -10                      -- but its a prominent absorption line
  and Q.ObjID = N.ObjID               -- N is a neighbor record
  and G.ObjID = N.NeighborObjID       -- G is a neighbor of Q
  and N.distanceMins < (10.0/60.0)    -- and it is within 10 arcseconds of the Q.
```

The query finds 975 objects in 5 seconds elapsed and 4 seconds of CPU time. The diagram below shows that the optimizer first parallel hash joins the SpecObj table with the QSO class, and then streams the result to a parallel hash join of the SpecLine table. Any tuples that qualify are then nested loops joined to the Neighbors table to find qualifying neighbors (within 10 arcseconds). Qualifying objects are then joined with the photoObj table to see if they are primary galaxies. If so, the QSO and galaxy pair is output.

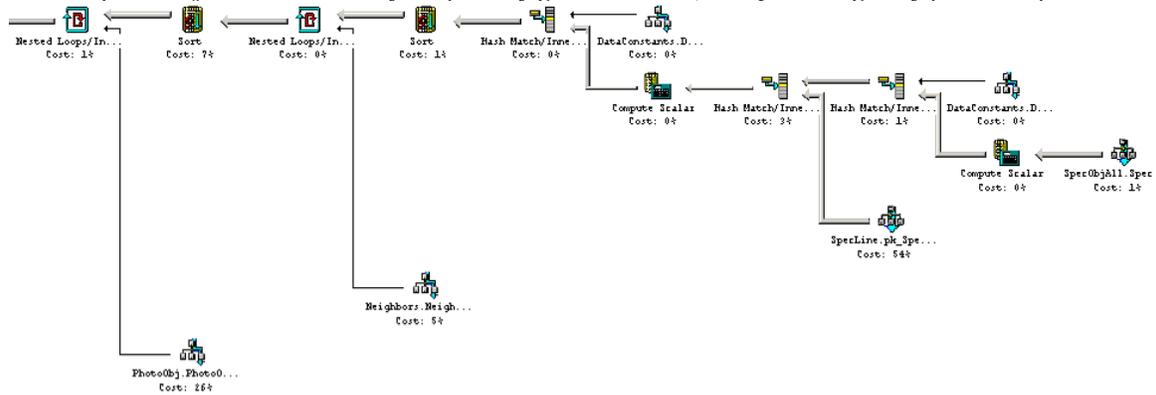

**Q20: For each galaxy in the LRG data set (Luminous Red Galaxy), in 160<right ascension<170, count of galaxies within 30"of it that have a photoZ within 0.05 of that galaxy.**

First form the LRG (Luminous Red Galaxy) table. Then scan for galaxies in clusters (the cluster is their parent object) with a predicate to limit the region of the sky. For each galaxy, test with a sub-query that no other galaxy in the same cluster is brighter. Then do a spatial join of this table with the galaxies to return the desired counts. For a galaxy in the BCG data set (brightest color galaxy), in 160<right ascension<170, 25<declination<35, give a count of galaxies within 30" which have a photoZ within 0.05 of the BCG.

```sql
declare @binned       bigint;                                -- initialized "binned" literal
set     @binned =     dbo.fPhotoFlags('BINNED1') +           -- avoids SQL2K optimizer problem
                      dbo.fPhotoFlags('BINNED2') +
                      dbo.fPhotoFlags('BINNED4') ;
declare @blended      bigint;                                -- initialized "blended" literal
set     @blended =    dbo.fPhotoFlags('BLENDED');            -- avoids SQL2K optimizer problem
declare @noDeBlend    bigint;                                -- initialized "noDeBlend" literal
set     @noDeBlend =  dbo.fPhotoFlags('NODEBLEND');          -- avoids SQL2K optimizer problem
declare @child        bigint;                                -- initialized "child" literal
set     @child =      dbo.fPhotoFlags('CHILD');              -- avoids SQL2K optimizer problem
declare @edge         bigint;                                -- initialized "edge" literal
set     @edge =       dbo.fPhotoFlags('EDGE');               -- avoids SQL2K optimizer problem
declare @saturated    bigint;                                -- initialized "saturated" literal
set     @saturated=   dbo.fPhotoFlags('SATURATED');          -- avoids SQL2K optimizer problem
select  G.objID, count(*) as pop
into    ##results
from    Galaxy     as G,                    -- first gravitational lens candidate
        Neighbors  as N,                    -- precomputed list of neighbors
        Galaxy     as U,                    -- a neighbor galaxy of G
        PhotoZ     as GpZ,                  -- photoZ of first galaxy
        PhotoZ     as NpZ                   -- photoZ of second galaxy
where   G.objID = N.objID                   -- connect G and U via the neighbors table
   and  U.objID = N.neighborObjID           -- so that we know G and U are within
   and  N.objID < N.neighborObjID           -- 30 arcseconds of one another.
   and  G.objID = GpZ.objID                 -- join to photoZ of G
   and  U.objID = NpZ.objID                 -- join to photoZ of N
   and  G.ra between 160 and 170            -- restrict search to a part of the sky
   and  G.dec between -5 and 5              -- that is in database
   and  abs(GpZ.Z - NpZ.Z) < 0.05           -- restrict the photoZ differences
   -- Color cut for an BCG courtesy of James Annis of Fermilab
   and (G.flags & @binned) > 0
   and (G.flags & ( @blended + @noDeBlend + @child)) != @blended
   and (G.flags & (@edge + @saturated)) = 0
   and  G.petroMag_i > 17.5
   and (G.petroMag_r > 15.5 or G.petroR50_r > 2)
   and (G.g >0 and G.r >0 and G.i >0)
   and (  (   ((G.petroMag_r-G.reddening_r)    < 19.2)
          and ((G.petroMag_r - G.reddening_r)
                             < (12.38 + (7/3)*( G.g- G.r ) + 4 *( G.r - G.i ) ) )
          and ((abs( G.r - G.i - (G.g - G.r )/4 - 0.18 )) < 0.2)
          and ((G.petroMag_r - G.reddening_r +
                             2.5*Log10(2*pi()*G.petroR50_r* G.petroR50_r )) < 24.2  )
          )
       or (   ((G.petroMag_r - G.reddening_r)       < 19.5                         )
          and ((G.r - G.i - (G.g - G.r)/4 - 0.18 ) > (0.45 - 4*( G.g- G.r ) )    )
          and ((G.g - G.r ) > ( 1.35 + 0.25 *( G.r - G.i ) )                       )
          and ((G.petroMag_r - G.reddening_r  +
                             2.5*Log10(2*pi()*G.petroR50_r* G.petroR50_r )) < 23.3  )
        ) )
group by G.objID
```

The query returns 690 objects in 355 seconds elapsed and 16 seconds of CPU time. This query nested loops joins the qualifying BCG galaxies with their neighbors table to get a list of neighbor object IDs. It then nested loops joins with the *PhotoZ* to get the *PhotoZ* of the BCG. It then picks up the neighbor's details, and if that qualifies, it picks up the *photoZ* of the neighbor, all using parallel nested loops joins.

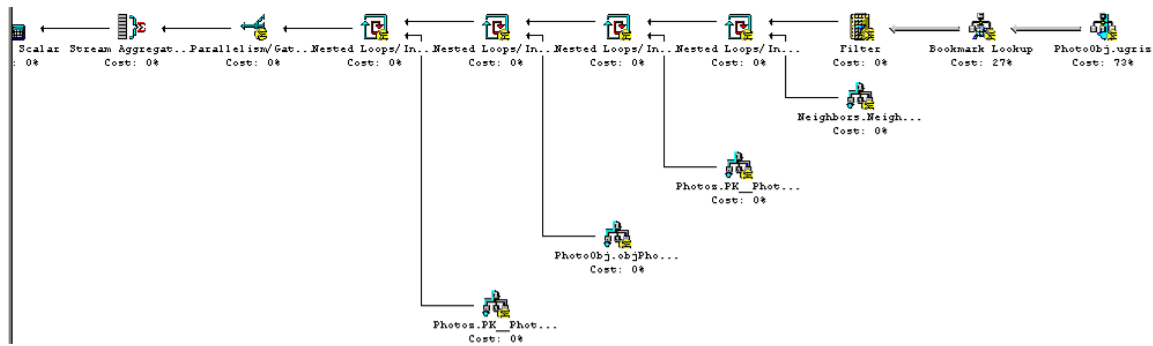

| Query | # | CPU time (s) | Elapsed Time (s) | IO count | Records returned | Comment |
|---|---|---|---|---|---|---|
| Find all galaxies without saturated pixels within 1' of a given point. | 1 | 0.05 | 0.19 | 39 | 19 | Spatial index lookup, then join with photoObj table. |
| Find all galaxies with blue surface brightness between and 23 and 25 magnitude per square arcseconds, and -10<super galactic latitude (sgb) between (-10º. 10º), and declination less than zero. | 2 | 14.1 | 18.63 | 28,740 | 191,062 | Sequential-parallel scan of an index   Much of the time goes into inserting 191k records into the results table.  If one just counts the records it is 2 seconds less CPU time and 10 seconds less elapsed time. |
| Find all galaxies brighter than magnitude 22, where the local extinction is >0.75. | 3 | 512.1 | 248.90 | 1,994,001 | 488,183 | A parallel sequential scan of the ugriz index and then a bookmark lookup, and then and then inserting the answers in a temporary table . |
| Find galaxies with an isophotal surface brightness (SB) larger than 24 in the red band, with an ellipticity>0.5, and with the major axis of the ellipse between 30" and 60"arc seconds (a large galaxy). | 4 | 8.5 | 17.77 | 26,311 | 787 | Parallel scan of the NEO index (covers type, status, flags, isoA_r, q_r, and u_r) then does a bookmark lookup on base table for rho. |
| Find all galaxies with a deVaucouleours profile (r¼ falloff of intensity on disk) and the photometric colors consistent with an elliptical galaxy. | 5 | 65.8 | 166.10 | 406,377 | 14,544 | Parallel table scan of PhotoObj with a fairly complex predicate. |
| Find galaxies that are blended with a star and output the deblended galaxy magnitudes. | 6 | 49.6 | 40.50 | 34,678 | 1,088,806 | Scans parent index of photoObj looking for galaxies, joins with parent's children to see if they are stars/ |
| Provide a list of star-like objects that are 1% rare | 7 | 90.2 | 53.39 | 21,197 | 476 | A two-step query.  First builds a hash-aggregate on magnitude- differences using the UGRIZ index.  Then deletes the most popular buckets from the answer set. |
| Find all objects with unclassified spectra. | 8 | 0.03 | 0.13 | 85 | 260 | A scab if the SpecObj table. |
| Find quasars with a line width >2000 km/s and 2.5<redshift<2.7. | 9 | 0.34 | 0.44 | 316 | 54 | A sequential scan of the spectra table and a nested-loops join with the SpecLine table. |
| Find galaxies with spectra that have an equivalent width in Ha >40Å | 10 | 4.8 | 4.64 | 9,441 | 5,496 | A nested loops join of a  sorted hash join of a nested loops join. |
| Find galaxies with spectra that have an equivalent width in Ha >40Å  and Ha/Hb > .2 | 10A | 1.9 | 1.31 | 53 | 9 | Two nested loops joins of a sorted pair of nested loops joins. |
| Find all elliptical galaxies with spectra that have an anomalous emission line. | 11 | 32.8 | 28.03 | 28,413 | 22,086 | A hash join of a pair of nested-loops joins which in turn consume nested loops joins. 6 joins and a sort in all. |
| Create a grided count of galaxies with u-g>1 and r<21.5 over -5<declination<5, and 175<right ascension<185, on a grid of 2' arc minutes.  Create a map of masks over the same grid. | 12 | 3.7 | 4.94 | 2,002 | 26,669 + 135 | Does a spatial join with an HTM area to subset the galaxies, then builds a hash-table counting the populations.  Does a second spatial join to count the mask cells. |
| Create a count of galaxies for each of the HTM triangles which satisfy a certain color cut, like 0.7u-0.5g-0.2i<1.25  and r<21.75, output it in a form adequate for visualization. | 13 | 28.1 | 20.13 | 28,404 | 7,604 | A hash aggregate based on a scan of the xyz index. |
| Find stars with multiple measurements that have magnitude variations >0.1. | 14 | 107.73 | 118.05 | 116,328 | 48,245 | This is a parallel merge join feeding a parallel hash join. |
| Provide a list of moving objects consistent with an asteroid. | 15A | 72.2 | 162 | 406,385 | 1,303 | A simple table scan computing the velocities and selecting those objects that have high velocity. |
| Find fast-moving near-earth objects. | 15B | 50.4 | 55.44 | 20,958 | 4 | A nested loops join of the NEO index with itself. |

| Query | # | CPU time (s) | Elapsed Time (s) | IO count | Records returned | Comment |
|---|---|---|---|---|---|---|
| Find all objects similar to the colors of a quasar at 5.5<redshift<6.5. | 16 | 14.9 | 13.61 | 21131 | 1 | Parallel scan of the UGRIZ index applying the color cut predicate. |
| Find binary stars where at least one of them has the colors of a white dwarf. | 17 | 70.8 | 60.91 | 149,219 | 2,775 | Parallel scan UGRIZ index with a color-cut predicate to get candidates. Parallel nested join that with the Neighbors and then nested join with PhotoObj to get the answer. |
| Find all objects within 30 arcseconds of one another that have very similar colors: that is where the color ratios u-g, g-r, r-i are less than 0.05m. | 18 | 1475 | 929.89 | 562,320 | 4,442 | Parallel merge join of a PhotoObj with Sort of parallel nested loops join of PhotoObj with Neighbors. |
| Find quasars with a broad absorption line in their spectra and at least one galaxy within 10 arcseconds. Return both the quasars and the galaxies. | 19 | 3.9 | 4.67 | 8,070 | 975 | Nested loops join of a sort of a nested loops join of a hash join of a hash join of a hash join. |
| For each galaxy in the LRG data set (Luminous Red Galaxy), in 160<right ascension<170, count of galaxies within 30"of it that have a photoZ within 0.05 of that galaxy. | 20 | 355.7 | 497.7 | 2,326,775 | 690 | Four nested loops joins in series. |
| Find Cataclysmic variables. | SX1 | 23.27 | 18.56 | 28,745 | 358,545 | Scan of UGRIZ index. |
| Find high-velocity objects | SX2 | 64.7 | 170.81 | 408,357 | 239,996 | Parallel scan of PhotoObj table. |
| Find a spatial subset/ | SX3 | 56.1 | 168.42 | 408,503 | 988,441 | Parallel scan of PhotoObj table. |
| Lookup object by field,obj | SX4 | 1.7 | 3.15 | 4,286 | 24 | Key lookup. |
| Galaxies with blue centers | SX5 | 544 | 354.08 | 2,337,686 | 1,686,559 | Parallel scan of UGRIZ index |
| The PSF colors of all stars brighter than 20th (rejecting on various flags) that have PSP_STATUS = 2 | SX6 | 1.1 | 1.50 | 4,390 | 1,048 | Parallel nested loops join of PhotoObj table with Field table. |
| Find clusters | SX7 | 534.9 | 380.5 | 406,453 | 5,261,445 | Parallel scan of PhotoObj table (with a bookmark lookup) |
| Diameter-limited galaxy sample. | SX8 | 531.2 | 349.1 | 30,368 | 255,332 | Parallel scan of UGRIZ index (with a bookmark lookup) |
| Extremely red galaxies. | SX9 | 167.4 | 152.56 | 356,014 | 189,241 | Parallel nested loops join of field table with a parallel sort of a parallel scan of PhotoObj UGRIZ and a bookmark lookup. |
| The BRG sample | SX10 | 559 | 360.45 | 2,319,372 | 28,303 | A parallel sequential scan of PhotoObj table with a bookmark lookup |
| Low-z QSO candidates | SX11 | 13.2 | 20.7 | 21,270 | 41,935 | A sequential scan of the UGRIZ index. |
| Errors on moving objects | SX12 | 50.4 | 104.0 | 406,379 | 583,635 | A parallel sequential scan of PhotoObj table. |
| A random sample of the data. | SX13 | 49.7 | 170.28 | 406,658 | 59,536 | A parallel sequential scan of PhotoObj table. |
| Cross match. | SX14 | - | - | - | - | -- done by website |
| Find quasars | SX15 | 10.8 | 14.36 | 21,792 | 337 | A sequential scan of the UGRIZ index. Followed by bookmark lookup to base table to get additional attributes. |

```
--------------------------------------------------------------------------
--Query SX1: Cataclysmic variables
Paula Szkody <szkody@alicar.astro.washington.edu>
Search for Cataclysmic Variables and pre-CVs with White Dwarfs and very late secondaries:
        u-g < 0.4
        g-r < 0.7
        r-i > 0.4
        i-z > 0.4
--------------------------------------------------------------------------
SELECT  run, camCol, rerun, field, objID,
        u,g,r,i,z,
        ra,  dec
INTO    ##results
FROM    PhotoPrimary
WHERE   (u - g) < 0.4 and
        (g - r) < 0.7 and
        (r - i) > 0.4 and
        (i - z) > 0.4
--------------------------------------------------------------------------
```

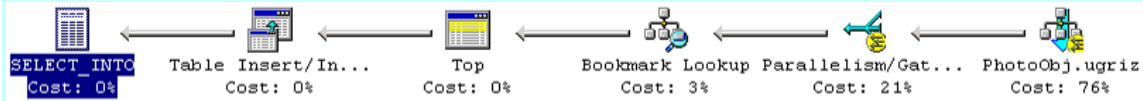

```
--------------------------------------------------------------------------
--Query SX2: Velocities and errors ===========================
(Robert H. Lupton <rhl@astro.princeton.edu>)
--------------------------------------------------------------------------
SELECT  run, camCol, field, objID,
        rowC, colC, rowV, colV, rowVErr, colVErr, flags,
        psfMag_u,psfMag_g,psfMag_r,psfMag_i,psfMag_z,
        psfMagErr_u,psfMagErr_g,psfMagErr_r,psfMagErr_i,psfMagErr_z
INTO ##results
FROM PhotoPrimary
WHERE   ((rowv * rowv) / (rowvErr * rowvErr) +
        (colv * colv) / (colvErr * colvErr) > 4)
--------------------------------------------------------------------------
```

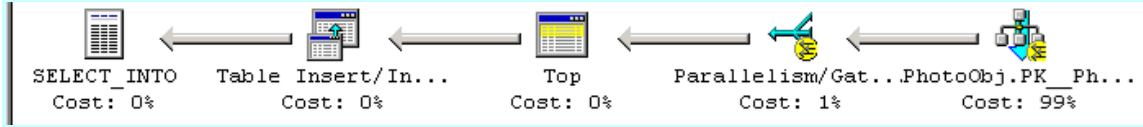

```
--------------------------------------------------------------------------
--Query SX3: Coordinate cut
(Robert H. Lupton <rhl@astro.princeton.edu>)
coordinate cut --> cut in ra --> 40:100

--------------------------------------------------------------------------
SELECT colc_g, colc_r
INTO ##results
FROM   PhotoObj
WHERE  (-0.642788 * cx + 0.766044 * cy>=0) AND
       (-0.984808 * cx - 0.173648 * cy <0)
--------------------------------------------------------------------------
```

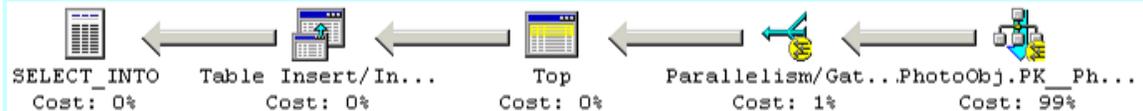

```
--------------------------------------------------------------------------
--Query SX4: Searching objects and fields by ID
(Robert H. Lupton <rhl@astro.princeton.edu>)
Searching for a particular object in a particular field.
--------------------------------------------------------------------------
SELECT *
INTO    ##results
FROM    PhotoObj
WHERE   obj  = 14 AND  field = 270
--------------------------------------------------------------------------
```

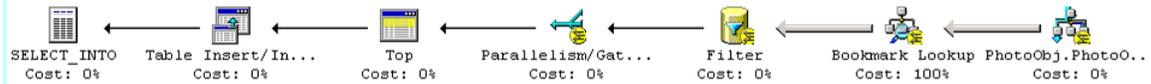

```
--------------------------------------------------------------------------
--Query SX5: Galaxies with bluer centers
Michael Strauss <strauss@astro.princeton.edu>
For all galaxies with r_Petro < 18, not saturated, not bright, not edge,
give me those whose centers are appreciably bluer than their outer parts.
That is, define the center color as: u_psf - g_psf
And define the outer color as: u_model - g_model
Give me all objects that have (u_model - g_model) - (u_psf - g_psf) < -0.4
--------------------------------------------------------------------------
DECLARE @flags  BIGINT;
SET @flags =   dbo.fPhotoFlags('SATURATED') +
               dbo.fPhotoFlags('BRIGHT')    +
               dbo.fPhotoFlags('EDGE')
SELECT colc_u, colc_g,  objID       --or whatever you want from each object
INTO   ##results
FROM    Galaxy
WHERE   (Flags &  @flags )= 0
        and petroRad_r < 18
        and ((colc_u - colc_g) - (psfMag_u - psfMag_g)) < -0.4
--------------------------------------------------------------------------
```

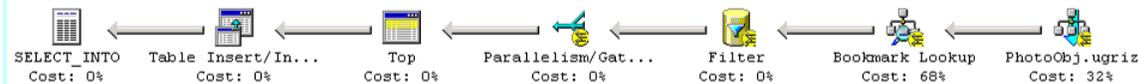

```
--------------------------------------------------------------------------
--Query SX6:  PSF colors of stars
Michael Strauss <strauss@astro.princeton.edu>
Give me the PSF colors of all stars brighter than 20th (rejecting on various flags) that
have PSP_STATUS = 2
--------------------------------------------------------------------------
SELECT  s.psfMag_g,          -- or whatever you want from each object
s.run,
s.camCol,
s.rerun,
s.field
INTO ##results
FROM Star s, Field f
WHERE  s.fieldID = f.fieldID
  AND  s.psfMag_g < 20
  AND  f.pspStatus = 2
--------------------------------------------------------------------------
```

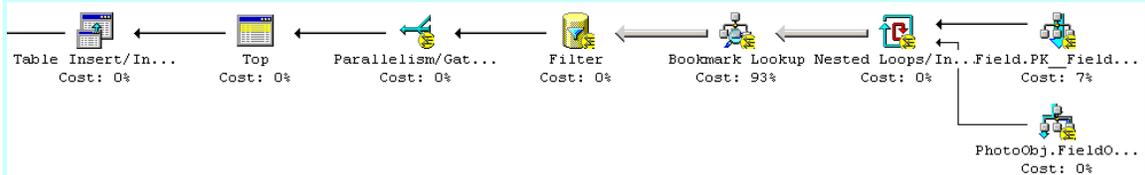

```
-----------------------------------------------------------------------
--Query SX7:   Cluster finding
(James Annis <annis@fnal.gov>)
if {AR_DFLAG_BINNED1 || AR_DFLAG_BINNED2 || AR_DFLAG_BINNED4} {
 if {! ( AR_DFLAG_BLENDED AND !( AR_DFLAG_NODEBLEND || AR_DFLAG_CHILD))} {
  if { galaxy } { ;# not star, asteroid, or bright
   if { primary_object} {
    if {petroMag{i} < 23 } { accept }
    }}}}
-----------------------------------------------------------------------
DECLARE @binned BIGINT
SET @binned =  dbo.fPhotoFlags('BINNED1') +
               dbo.fPhotoFlags('BINNED2') +
               dbo.fPhotoFlags('BINNED4')
DECLARE @deblendedChild BIGINT
SET @deblendedChild =  dbo.fPhotoFlags('BLENDED')   +
                       dbo.fPhotoFlags('NODEBLEND') +
                       dbo.fPhotoFlags('CHILD')
DECLARE @blended BIGINT
SET @blended = dbo.fPhotoFlags('BLENDED')
SELECT   camCol, run, rerun, field, objID, ra, dec
INTO ##results
FROM Galaxy                             -- select galaxy and primary only
WHERE (flags & @binned )> 0
  AND (flags & @deblendedChild ) !=  @blended
  AND petroMag_i < 23
-----------------------------------------------------------------------
```

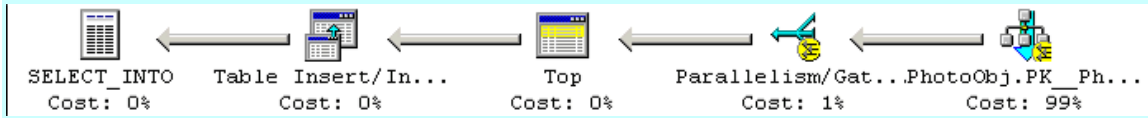

```
--------------------------------------------------------------------
--Query SX8: Diameter-limited sample of galaxies
(James Annis <annis@fnal.gov>)
if {AR_DFLAG_BINNED1 || AR_DFLAG_BINNED2 || AR_DFLAG_BINNED4} {
 if {! ( AR_DFLAG_BLENDED AND !( AR_DFLAG_NODEBLEND || AR_DFLAG_CHILD)} {
  if { galaxy } { ;# not star, asteroid, or bright
   if (!AR_DFLAG_NOPETRO) {
    if { petrorad > 15 } { accept }
      } else {
     if { petror50 > 7.5 } { accept }
     }
    if (AR_DFLAG_TOO_LARGE AND petrorad > 2.5 ) { accept }
    if ( AR_DFLAG_SATUR AND petrorad < 17.5) { don't accept }
 }}}
--------------------------------------------------------------------
DECLARE @binned BIGINT
SET @binned =  dbo.fPhotoFlags('BINNED1') +
               dbo.fPhotoFlags('BINNED2') +
               dbo.fPhotoFlags('BINNED4')
DECLARE @deblendedChild BIGINT
SET @deblendedChild = dbo.fPhotoFlags('BLENDED')   +
                      dbo.fPhotoFlags('NODEBLEND') +
                      dbo.fPhotoFlags('CHILD')
DECLARE @blended BIGINT
SET @blended = dbo.fPhotoFlags('BLENDED')
DECLARE @noPetro BIGINT
SET @noPetro = dbo.fPhotoFlags('NOPETRO')
DECLARE @tooLarge BIGINT
SET @tooLarge = dbo.fPhotoFlags('TOO_LARGE')
DECLARE @saturated BIGINT
SET @saturated = dbo.fPhotoFlags('SATURATED')
SELECT run, camCol, rerun, field, objID, ra, dec
INTO ##results
FROM Galaxy
WHERE (flags &  @binned )> 0
  AND (flags &  @deblendedChild ) !=  @blended
  AND (   ((flags & @noPetro    = 0) AND (petroRad_i > 15  ))
       OR ((flags & @noPetro    > 0) AND (petroRad_i > 7.5 ))
       OR ((flags & @tooLarge   > 0) AND (petroRad_i > 2.5 ))
       OR ((flags & @saturated  = 0) AND (petroRad_i > 17.5))
       )
--------------------------------------------------------------------
```

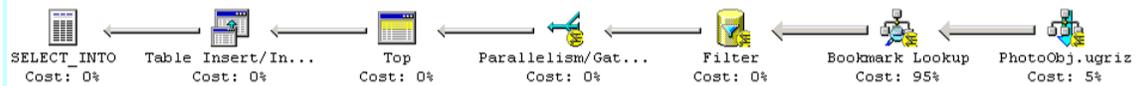

```
------------------------------------------------------------------------
--Query SX9: Extremely red galaxies:
(James Annis <annis@fnal.gov>)
if {AR_DFLAG_BINNED1 || AR_DFLAG_BINNED2 || AR_DFLAG_BINNED4} {
 if {! ( AR_DFLAG_BLENDED AND !( AR_DFLAG_NODEBLEND || AR_DFLAG_CHILD)} {
  if { galaxy } { ;# not star, asteroid, or bright
   if { primary_object} {
    if {!AR_DFLAG_CR AND !R_DFLAG_INTERP}
     if { frame_seeing < 1.5" } {
      if { Mag_model<i>-Mag_model<z> -
        (reddening<i> - reddening<z>) > 1.0 }
         { accept }
   }}}}}}
------------------------------------------------------------------------
DECLARE @binned BIGINT
SET @binned =  dbo.fPhotoFlags('BINNED1') +
               dbo.fPhotoFlags('BINNED2') +
               dbo.fPhotoFlags('BINNED4')
DECLARE @deblendedChild BIGINT
SET @deblendedChild =  dbo.fPhotoFlags('BLENDED')   +
                       dbo.fPhotoFlags('NODEBLEND') +
                       dbo.fPhotoFlags('CHILD')
DECLARE @blended BIGINT
SET @blended = dbo.fPhotoFlags('BLENDED')
DECLARE @crIntrp BIGINT
SET @crIntrp = dbo.fPhotoFlags('COSMIC_RAY')  +
               dbo.fPhotoFlags('INTERP')
SELECT g.run, g.camCol, g.rerun, g.field, g.objID, g.ra, g.dec
INTO ##results
FROM galaxy g, Field f
WHERE g.fieldID = f.fieldID
 AND (flags &  @binned )> 0
 AND (flags &  @deblendedChild ) !=  @blended
 AND (flags &  @crIntrp ) = 0
 AND f.psfWidth_r < 1.5
 AND (i - z) > 1.0
------------------------------------------------------------------------
```

Top — Cost: 0%  
Parallelism/Gat.. — Cost: 0%  
Nested Loops/In... — Cost: 0%  
Sort — Cost: 0%  
Filter — Cost: 0%  
Bookmark Lookup — Cost: 35%  
PhotoObj.ugriz — Cost: 65%  

Field.PK__Field... — Cost: 0%

```
-------------------------------------------------------------------
--Query SX10: The BRG sample
(James Annis <annis@fnal.gov>)
if {AR_DFLAG_BINNED1 || AR_DFLAG_BINNED2 || AR_DFLAG_BINNED4} {
   if {! ( AR_DFLAG_BLENDED AND !( AR_DFLAG_NODEBLEND || AR_DFLAG_CHILD)} {
      if {!AR_DFLAG_EDGE & !AR_DFLAG_SATUR} {
         if { galaxy} { ;# not star, asteroid, or bright
            if { primary_object} {
               if {! (petroMag<2> < 15.5 AND petror50<2> < 2) } {
                  if {petroMag<r> > 0 AND Mag_model<g> > 0 AND
                      Mag_model<r> > 0 AND Mag_model<i> > 0 } {
                           petSB = deRed_r + 2.5*log10(2*3.1415*petror50<r>^2)
                           deRed_g = petroMag<g> - reddening<g>
                           deRed_r = petroMag<r> - reddening<r>
                           deRed_i = petroMag<i> - reddening<i>
                           deRed_gr = deRed_g - deRed_r
                           deRed_ri = deRed_r - deRed_i
                           cperp = deRed_ri - deRed_gr/4.0 - 0.18
                           cpar = 0.7*deRed_gr + 1.2*(deRed_ri -0.18)
                  if {(deRed_r < 19.2 AND deRed_r < 13.1 + cpar/0.3 AND
                       abs(cperp) < 0.2 AND petSB < 24.2 ) ||
                      (deRed_r < 19.5 AND cperp > 0.45 - deRed_gr/0.25 AND
                       deRed_gr > 1.35 + deRed_ri*0.25 AND petSB < 23.3) {
                                accept ;# whew!!!
   } } } } }  } } } } }
-------------------------------------------------------------------
DECLARE @binned BIGINT
SET @binned =  dbo.fPhotoFlags('BINNED1') +
               dbo.fPhotoFlags('BINNED2') +
               dbo.fPhotoFlags('BINNED4')
DECLARE @deblendedChild BIGINT
SET @deblendedChild =  dbo.fPhotoFlags('BLENDED')   +
                       dbo.fPhotoFlags('NODEBLEND') +
                       dbo.fPhotoFlags('CHILD')
DECLARE @blended BIGINT
SET @blended = dbo.fPhotoFlags('BLENDED')
DECLARE @edgedSaturated BIGINT
SET @edgedSaturated =  dbo.fPhotoFlags('EDGE') +
                       dbo.fPhotoFlags('SATURATED')
SELECT  run, camCol, rerun, field, objID, ra, dec
INTO ##results
FROM Galaxy as G
WHERE (flags &  @binned)> 0
 and (flags &  @deblendedChild) != @blended
 and (flags &  @edgedSaturated) = 0
 and  G.petroMag_i > 17.5
 and (G.petroMag_r > 15.5 or G.petroR50_r > 2)
 and (G.g >0 and G.r >0 and G.i >0)
 and ( (   ((G.petroMag_r-G.reddening_r)    < 19.2)
           and ((G.petroMag_r - G.reddening_r)
                             < (12.38 + (7/3)*( G.g-  G.r ) + 4 *( G.r - G.i ) ) )
           and ((abs( G.r - G.i - (G.g - G.r )/4 - 0.18 )) < 0.2)
           and ((G.petroMag_r - G.reddening_r +
                       2.5*Log10(2*pi()*G.petroR50_r* G.petroR50_r )) < 24.2  )
        )
      or (   ((G.petroMag_r - G.reddening_r)        < 19.5                     )
           and ((G.r - G.i - (G.g - G.r)/4 - 0.18 ) > (0.45 - 4*( G.g- G.r ) )   )
           and ((G.g - G.r ) > ( 1.35 + 0.25 *( G.r - G.i ) )                  )
           and ((G.petroMag_r - G.reddening_r   +
                       2.5*Log10(2*pi()*G.petroR50_r* G.petroR50_r )) < 23.3  )
       ) )
```

[query plan diagram: SELECT_INTO → Table Insert/In... → Top → Filter → PhotoObj.PK_Ph...]

```
-------------------------------------------------------------------
--Query SX11:  Low-z QSO candidates
Gordon Richards <richards@oddjob.uchicago.edu>
Low-z QSO candidates using the following cuts:
--
-0.27 <= u-g < 0.71
```

```
-0.24 <= g-r < 0.35
-0.27 <= r-i < 0.57
-0.35 <= i-z < 0.70
g <= 22
objc_type == 3
declare @cpu int, @physical_io int, @clock datetime, @elapsed int;
exec dbo.InitTimeX  @clock OUTPUT, @cpu OUTPUT, @physical_io OUTPUT

----------------------------------------------------------------------
SELECT g, objID      -- or whatever you want returned
INTO ##results
FROM Galaxy          -- takes care of objc_type == 3
WHERE (g <= 22) AND
(u-g >= -0.27) AND (u-g < 0.71) AND
(g-r >= -0.24) AND (g-r < 0.35) AND
(r-i >= -0.27) AND (r-i < 0.57) AND
(i-z >= -0.35) AND (i-z < 0.70)
```

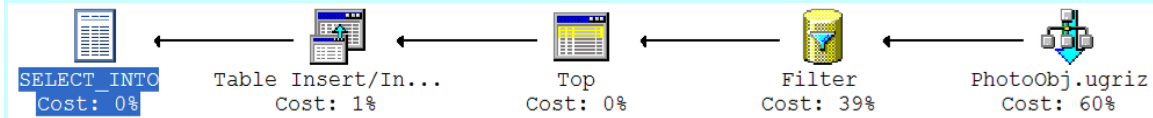

```
----------------------------------------------------------------------
--Query SX12: Errors on moving objects
Gordon Richards <richards@oddjob.uchicago.edu>
Another useful query is to see if the errors on moving (or apparently
moving objects) are correct. For example it used to be that some
known QSOs were being flagged as moving objects. One way to look for
such objects is to compare the velocity to the error in velocity and
see if the "OBJECT1_MOVED" or "OBJECT2_BAD_MOVING_FIT" is set. So
return objects with
--
objc_type == 3
sqrt(rowv*rowv + colv*colv) >= sqrt(rowvErr*rowvErr + colvErr*colvErr)
--
then output, the velocity, velocity errors, i' magnitude, and the
relevant "MOVING" object flags.
----------------------------------------------------------------------
DECLARE @moved BIGINT
SET     @moved = dbo.fPhotoFlags('MOVED')
DECLARE @badMove BIGINT
SET     @badMove = dbo.fPhotoFlags('BAD_MOVING_FIT')
SELECT rowv, colv, rowvErr, colvErr, i,
       (flags & @moved)    as MOVED,
       (flags & @badMove)  as BAD_MOVING_FIT
INTO ##results
FROM Galaxy
WHERE   (flags & (@moved + @badMove)) > 0
AND (rowv * rowv + colv * colv) >=
                    (rowvErr * rowvErr + colvErr * colvErr)
```

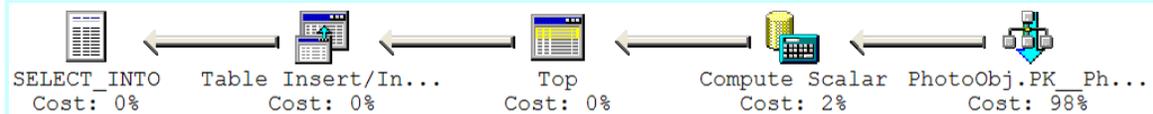

```
----------------------------------------------------------------------
--Query SX13: A random sample of the data
Karl Glazebrook <kgb@pha.jhu.edu>
So as a newcomer I might want to do something like 'give me the colours
of 100,000 random objects from all fields which are survey quality' so
then I could plot up colour-colour diagrams and play around with
more sophisticated cuts. How would I do that?
declare @cpu int, @physical_io int, @clock datetime, @elapsed int;
exec dbo.InitTimeX  @clock OUTPUT, @cpu OUTPUT, @physical_io OUTPUT
----------------------------------------------------------------------
SELECT u,g,r,i,z
INTO   ##results
FROM   Galaxy
```

```
WHERE (obj  % 100 )= 1
```

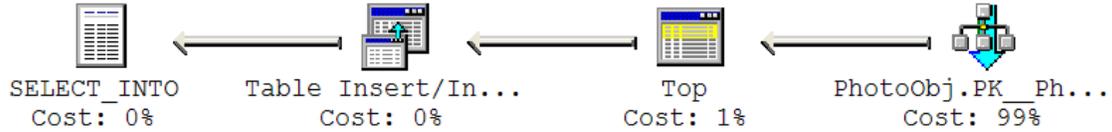

```
--------------------------------------------------------------------
--Query SX14:--   (Dan VandenBerk <danvb@fnal.gov>)
We have a list of objects -- RA and DEC -- for which we would like
matches and all of the Obj data. Can I send you the list?

---------------------------------------------------------------------
Done as a download from website: see http://skyserver.sdss.org/v3/en/tools/crossid/

---------------------------------------------------------------------

--===================================================================

---------------------------------------------------------------------
--Query SX15:  Find quasars
(Xiaohui Fan et.al. <fan@astro.princeton.edu>)
---------------------------------------------------------------------
SELECT run, camCol, rerun, field, objID,
       u,g,r,i,z,
       ra, dec
INTO   ##results
FROM   Star         -- or Galaxy
WHERE  (((u – g) > 2.0) or (u > 22.3 ) )
  AND ( i < 19 )
  AND ( i > 0 )
  AND ((g – r) > 1.0 )
  AND (((r – i) < (0.08 + 0.42 * (g – r – 0.96))) OR ((g – r) > 2.26 ) )
  AND ((i – z)  < 0.25 )
```

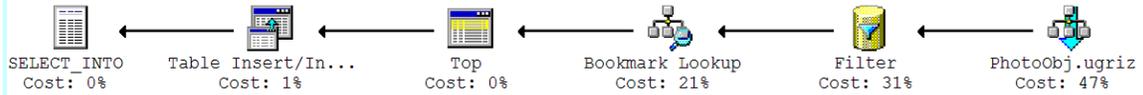